\documentclass{aa}
\usepackage{graphicx}
\usepackage{txfonts}
\usepackage{booktabs}

\def\d{\displaystyle}
\def\kms{km\,s$^{-1}$}

\def\vs{$v\sin{i}$}
\def\te{$T_{\rm eff}$}
\def\lg{$\log{g}$}
\def\vt{$v_{\rm turb}$}

\def\cd{c\,d$^{-1}$}

\begin{document}

\title{Spectroscopic long-term monitoring of RZ\,Cas\\ Part I: Basic stellar and system parameters
\thanks{Based on observations made with the Alfred Jensch Telescope at the Th\"uringer Landessternwarte Tautenburg and the Mercator Telescope, operated on the island of La Palma by the Flemish Community, at the Spanish Observatorio del Roque de los Muchachos of the Instituto de Astrof\'{i}sica de Canarias. }}

\author{H. Lehmann\inst{1}  \and A. Dervi\c so\u glu\inst{2,3} 
\and D.E. Mkrtichian\inst{4} \and F. Pertermann\inst{1}  \and
A. Tkachenko\inst{5} \and V. Tsymbal\inst{6}}

\institute{Th\"uringer Landesternwarte Tautenburg, Sternwarte 5, D-07778 Tautenburg, Germany
\and Department of Astronomy and Space Sciences, Erciyes University, 38039, Kayseri, Turkey
\and Astronomy and Space Sciences Observatory and Research Center, Erciyes University, 38039, Kayseri, Turkey
\and National Astronomical Research Institute of Thailand, 260 Moo 4, T. Donkaew, A. Maerim, Chiangmai, 50180, Thailand
\and Institute of Astronomy, KU Leuven, Celestijnenlaan 200D, B-3001 Leuven, Belgium
\and Institute of Astronomy, Russian Academy of Sciences, 119017, Pyatnitskaya Str. 48, Moscow, Russia }

\abstract{\object{RZ Cas} is a short-period Algol-type system showing episodes of mass transfer and $\delta$ Sct-like oscillations of its mass-gaining primary component. This system exhibits temporal changes in orbital period, \vs, and the oscillation pattern of the primary component.}
{We analyse high-resolution spectra of RZ\,Cas that we obtained during a spectroscopic long-term monitoring lasting from 2001 to 2017. In this first part we investigate the atmospheric parameters of the stellar components and  the time variation of orbital period, \vs, and radial velocities (RVs), searching for seasonal changes that could be related to episodes of mass exchange and to a possible activity cycle of the system triggered by the magnetic cycle of the cool companion.}
{We used spectrum synthesis to analyse the spectra of both components of RZ\,Cas. The study of variations of the orbital period is based on published times of primary minima.  We used the LSDbinary program to derive separated RVs and least-squares deconvolved (LSD) profiles of the components. From the LSD profiles of the primary we determined its \vs. Using Markov Chain Monte Carlo simulations with the PHOEBE program, we modelled the RV variations of both components.}
{Spectrum analysis resulted in precise atmospheric parameters of both components, in particular in surface abundances below solar values. We find that the variation of orbital period is semi-regular and derive different characteristic timescales for different epochs of observation. We show that the RV variations with orbital phase can be modelled when including two cool spots on the surface of the secondary component. The modelling leads to very precise masses and separation of the components. The seasonal variation of several parameters, such as \vs, rotation-orbit synchronisation factor, strength of the spots on the cool companion, and orbital period, can be characterised by a common timescale of the order of nine years.} 
{We interpret the timescale of nine years as the magnetic activity cycle of the cool companion. In particular the behaviour of the dark spots on the cool companion leads us to the interpretation that this timescale is based on an 18-year magnetic dynamo cycle. We conclude that the mass-transfer rate is controlled by the variable depth of the Wilson depression in the magnetic spot around the Lagrangian point L1. In the result, based on available data, we observe a damped activity cycle of the star, starting with a high mass-transfer episode around 2001 with a calculated mass-transfer rate of $1.5\times 10^{-6}$\,M$_\odot\,$yr$^{-1}$, followed by quiet periods in 2006 and 2009, slightly higher activity around 2013 and 2014, and again followed by quiet periods in 2015 and 2016. However, owing to missing data for years 2010 and 2011, we cannot exclude that a second high mass-transfer episode occurred within this time span.
}


\maketitle

\section{Introduction}

Various types of oscillating stars reside in binary systems and   therefore a precise determination of their system and stellar parameters can be performed. In particular the combined photometric-spectroscopic analysis of eclipsing binaries (EBs hereafter) leads to the determination of absolute masses and radii of the components in a direct way \citep[e.g.][]{2013A&A...557A..79L, 2013A&A...556A.138F}. Thanks to spectral disentangling techniques \citep[e.g.][]{1994A&A...281..286S, 1995A&AS..114..393H, 2001LNP...573..269I, 2019A&A...623A..31S}, faint components giving rise to a contribution of a few per cent can be detected in high-resolution spectra \citep[e.g.][]{2013A&A...557A..79L, 2014MNRAS.438.3093T, 2014MNRAS.443.3068B}. Such techniques ensure the determination of precise (up to one per cent) model-independent dynamical masses of stars, which can further be confronted with asteroseismic values provided at least one of the binary components pulsates. This also means that the results can be used to calibrate the asteroseismic mass determination that becomes more and more important, in particular for planet hosting stars \citep[e.g.][]{2007ApJ...670L..37H, 2008JPhCS.118a2016H, 2012A&A...543A..98H}.

Algol-type systems (Algols, hereafter) are semi-detached, interacting EBs consisting of a main-sequence star of spectral type B-A (primary, hereafter) and an evolved F-K type companion (secondary, hereafter). As a consequence of the above-mentioned facts, studying pulsating Algols from complementary spectroscopic and photometric data provides a valuable test of stellar evolutionary models. The group of oscillating eclipsing Algol stars (oEAs, hereafter) \citep{2002ASPC..259...96M} consists of eclipsing Algols with mass transfer where the mass-accreting primary shows $\delta$\,Sct-like oscillations. These stars are extraordinary objects for asteroseismic studies because they allow us to investigate short-term dynamical stellar evolution during mass-transfer episodes, most probably caused by the magnetic activity cycle of the less massive secondary. Basic principles of the interaction between the magnetic cycle of the cool secondary, the occurrence of rapid mass-transfer episodes, the dynamical behaviour of the system, and the excitation of different non-radial pulsation (NRP hereafter) modes of the mass-gaining primary can be studied in great detail.  \citet{2018MNRAS.475.4745M} showed for the first time that mass transfer and accretion influences amplitudes and frequencies of NRP modes of the oEA star RZ\,Cas, where amplitude changes are caused by the sensitivity of the mode selection mechanism to conditions in the outer envelope \citep[e.g.][]{1998A&A...333..141P}  and frequency variations by the acceleration of the outer layers by mass and angular momentum transfer.

The details of angular momentum exchange in Algols determining their evolution are still not completely understood. This is in particular valid for systems like the so-called R\,CMa stars  \citep[e.g.][]{2011MNRAS.418.1764B, 2013A&A...557A..79L, 2018A&A...615A.131L} that include companions of extremely low masses. But also for the oEA star RZ\,Cas, calculations showed that its actual configuration cannot be explained when assuming a purely conservative mass-transfer scenario in the past \citep{2008A&A...486..919M}. In several Algols, such as RZ\,Cas \citep{2008A&A...480..247L} or TW Dra \citep{2008A&A...489..321Z, 2010AJ....139.1327T}, orbital period variations were observed that can be attributed to periods of rapid mass exchange. The gas stream hits the equatorial zones of the atmosphere of the pulsating star, transfers an essential amount of angular momentum, and forces the acceleration of its outermost surface layers, thus causing strong differential rotation. While rotation alters the frequencies of the individual axisymmetric modes \citep[e.g.][]{1981ApJ...244..299S, 2008ApJ...679.1499L}, it also lifts the degeneracy in frequency for the non-axisymmetric modes as observed for some $\beta$\,Cep variables \citep[e.g.][]{2004A&A...415..241A, 2005MNRAS.360..619J}. This means that NRP modes are sensitive to the acceleration of the surface layers and that it is possible to probe the acceleration via the rotational splitting effect, as suggested by \citet{2018MNRAS.475.4745M} for RZ\,Cas. The study of the corresponding frequency shifts, together with the direct measurement of changes in the projected equatorial velocity (\vs, hereafter) of the primary from its spectral line profiles leads to an estimation of the amount of matter transferred to the primary. The results can be compared to those obtained from 3D hydrodynamical calculations, assuming different rates of mass transfer \citep[e.g.][]{2007ASPC..370..194M}, and finally explain the kind of mass and angular momentum transfer of the Algols.

One further advantage of oEAs is that the so-called spatial filtration or eclipse mapping effect, occurring as a result of the obscuration of parts of the stellar disc of the oscillating primary by the secondary during eclipses, simplifies the identification of the pulsation modes. The effect was predicted and found in photometric \citep[e.g.][]{2003ASPC..292..369G, 2004A&A...419.1015M, 2005ApJ...634..602R} and spectroscopic \citep{2018A&A...615A.131L} observations. The dynamic eclipse mapping method was introduced by \citet{2011MNRAS.416.1601B} with the aim of NRP mode identification by reconstructing the surface intensity patterns on EBs. With modern methods such as the least-squares deconvolution (LSD) technique \citep{1997MNRAS.291..658D}  and the pixel-by-pixel method of the FAMIAS program \citep{2008CoAst.157..387Z}, high-$l$-degree NRP modes can also be detected and identified from the signal-to-noise ratio (S/N hereafter) enhanced line profiles; a unique identification however is difficult \citep[e.g.][]{2009CoAst.159...45L}. 

RZ\,Cas (spectral type A3\,V\,+\,K0\,IV) is a short-period ($P$\,=\,1.1953\,d) Algol and one of the best studied oEA stars.  A partial eclipse is observed during primary minimum \citep{1994AJ....107.1141N}. The primary was found by \citet{1998IBVS.4581....1O, 2001AJ....122..418O} to exhibit short-period light variability with a dominant oscillation mode of a frequency of 64.2\,\cd. This finding was later confirmed from dedicated multi-site photometric campaigns by both \citet{2003ASPC..292..113M} and \citet{2004MNRAS.347.1317R}. The latter authors obtained simultaneous Str\"omgren uvby light curves of RZ\,Cas. These authors present a detailed photometric analysis for both binarity and pulsation, deriving WD \citep{1971ApJ...166..605W} solutions in the four mentioned passbands as well as absolute parameters of both components, and confirming the pulsational behaviour of the primary component as found by \citet{1998IBVS.4581....1O}. We use the radii and flux ratios between the components derived by  \citet{2004MNRAS.347.1317R} as a reference in our spectroscopic investigation. A comprehensive overview on the observations and analysis of RZ\,Cas can be found in \citet{2018MNRAS.475.4745M}. 

Starting our  spectroscopic investigation of RZ\,Cas in 2001, we were the first to detect rapid oscillations in its spectra \citep{2004A&A...413..293L}, and later on in spectra taken in 2006 \citep{2008A&A...480..247L, 2009A&A...504..991T}. From the different amplitudes of the Rossiter-McLaughlin effect \citep[][RME hereafter]{1924ApJ....60...15R, 1924ApJ....60...22M} observed during the primary eclipses in different seasons and the modelling of line profiles over the full orbital cycle using the Shellspec07\_inverse program, we deduced that RZ\,Cas was in an active phase of mass transfer in 2001, whereas in 2006 it was in a quiet state. To model the surface intensity distribution of the secondary of RZ\,Cas, we had to include a large cool spot facing the primary, presumably originating from a cooling mechanism by the enthalpy transport via the inner Lagrangian point as suggested by \citet{1994PASJ...46..613U}. Comparing the rapid oscillations found in the radial velocities (RVs hereafter) from 2001 and 2006 also with those derived from the light curves of RZ\,Cas taken over many years \citep[see][]{2018MNRAS.475.4745M}, we found that the NRP pattern of RZ\,Cas changed from season to season. Different NRP modes have been excited with different amplitudes in different years and also frequency variations of single modes were observed.  \citet{2018MNRAS.475.4745M} suggested that these frequency variations could be caused by a temporary acceleration of the outer layers of the primary owing to angular momentum exchange by mass-transfer effects.  

After observing RZ\,Cas in 2001 and 2006, we took new time series of high-resolution spectra in 2008 and 2009. The fact that we found a typical timescale of about nine years from the behaviour of the pulsation amplitudes but also from light-curve analysis \citep[see][]{2018MNRAS.475.4745M} forced us to start a spectroscopic monitoring of the star covering the years 2013 to 2017. We now investigate the complete data set with the aim to use the spectra and  observed variations in RVs and line profile variations (LPV hereafter) to deduce stellar and system parameters, check for timely variable NRP pulsation patterns, and try to correlate all observed variations with the occurrence of quiet and active phases of the Algol system.  Observations are described in Sect.\,\ref{Sect2}. The spectra taken with the HERMES spectrograph are used for a detailed analysis with the aim to derive precise atmospheric parameters for both components of RZ\,Cas (Sect.\,\ref{Sect3}). The extraction of RVs and calculation of mean, high-S/N line profiles with the newly developed LSDbinary program is described in Sect.\,\ref{Sect4}. Using different methods, we measure the projected equatorial rotation velocity of the primary (Sect.\,\ref{Sect5}). In Sect.\,\ref{Sect6} we use the O-C values collected from literature to compute the orbital period variations of RZ\,Cas over the last decades. We check for non-Keplerian effects in the orbital RV curves in Sect.\,\ref{Sect7.2} and try to model them using the PHOEBE \citep{2005ApJ...628..426P} program. The results are discussed in Sect.\,\ref{Sect8} followed by concluding remarks in Sect.\,\ref{Sect9}.

Our investigation of NRP is based on high-frequency oscillations in the RVs and in LPV. Applied methods and results will be described in a forthcoming article (Paper\,II) and discussed together with the results presented in this work.

\begin{table} \centering
 \tabcolsep 1.8mm
 \caption{Journal of observations listing the instrument, its spectral resolving power, year of observation, and mean Julian date. The last four columns give the number of spectra, total time span of observations in days, number of observed nights, and number of groups of observations.}\label{Tab01}
 \begin{tabular}{lclcrrrr}
 \toprule
 Source & $R$     & Year & mean JD     & $s$ & $t$ & $n$ & $g$\\
 \midrule
 TCES   & 32\,000 & 2001     & 2\,452\,190 &  962 &   15 &13&1\\
 TCES   & 32\,000 & 2006     & 2\,453\,800 &  517 &  150 &7 &3\\
 TCES   & 32\,000 & 2008     & 2\,454\,717 &   94 &   27 &5 &2\\
 HERMES & 85\,000 & 2009     & 2\,455\,156 &  228 &    5 &5 &1\\
 TCES   & 32\,000 & 2013     & 2\,456\,600 &  835 &  157 &21&3\\
 TCES   & 32\,000 & 2014     & 2\,456\,938 &  696 &   13 &8 &2\\ 
 TCES   & 58\,000 & 2015     & 2\,457\,300 &  998 &  152 &31&4\\
 TCES   & 58\,000 & 2016     & 2\,457\,647 &  586 &   14 &10&1\\
 TCES   & 58\,000 & 2017     & 2\,458\,097 & ~~43 &    4 &3 &1\\
 \bottomrule
 \end{tabular}
\end{table}

\section{Observations}\label{Sect2}

Spectra were taken over a total time span of 16 years with the TCES spectrograph at the 2 m Alfred Jensch Telescope of the Th\"uringer Landessternwarte (TLS) Tautenburg and over one year with the HERMES spectrograph \citep{2011A&A...526A..69R} at the 1.25 m Mercator Telescope on La Palma. The TCES instrument is an echelle spectrograph in Coude focus and covers the wavelength range 4450-7550\,\AA. A 2015 upgrade resulted in improved efficiency so that this spectrograph could be used with higher spectral resolution (narrower entrance slit of one instead of two arcsec width). Table\,\ref{Tab01} gives the journal of observations. 

The TCES spectra were reduced using standard MIDAS packages for Echelle spectrum reduction and HERMES spectra were reduced using the HERMES spectrum reduction pipeline. We used our own routines for the normalisation of the spectra to the local continuum. Instrumental shifts  were corrected using an additional calibration based on a large number of telluric O$_2$ absorption lines.

\section{Spectrum analysis}\label{Sect3}

\subsection{Spectra}

The HERMES spectra taken in  2009 comprise the  H$_\epsilon$ to H$_\alpha$ range, whereas the TLS spectra only include H$_\beta$ and H$_\alpha$. Moreover, the HERMES spectra provide the higher spectral resolution. That is why we decided to use them for spectrum analysis. We were able to decompose the spectra into the spectra of the components using the KOREL program \citep{1995A&AS..114..393H}. Analysing the decomposed spectra, we faced problems with the continuum normalisation. As a result of the very faint contribution of the secondary, small deviations in the continuum of the composite spectrum from the true continuum leads to large deviations in the continuum of the decomposed spectrum of the secondary. A first normalisation was applied during spectrum reduction by fitting higher polynomials to nodes assumed to represent the local continuum in each of the extracted echelle orders. This approach is not accurate enough when the contribution of the secondary is as faint as in the present case. Corrections to the local continuum have to be applied during spectrum analysis by a comparison with the continua of synthetic spectra. The spectra are then renormalised by multiplying them with the local ratio of the continua of the synthetic spectrum to that of the reduced spectrum. The spectrum decomposition with KOREL, on the other hand, is based on Fourier transform and introduces undulations in the continuum that are additive and have to be subtracted. This introduces an ambiguity with the multiplicative corrections described before that we could not solve to reach the required accuracy. Instead, we used the average of nine composite HERMES spectra of RZ\,Cas taken at maximum separation of the components and having about the same RV. In this case, we have to apply multiplicative continuum corrections only, for which we used spline functions. Because the signal of the secondary gets fainter for smaller wavelengths and because of the occurrence of telluric lines in the red part of the spectra we limited the spectral range to 4000-5550\,\AA, including the three Balmer lines H$_\beta$, H$_\gamma$, and H$_\delta$. 

\subsection{Methods}

\begin{figure}
 \includegraphics[width=\linewidth]{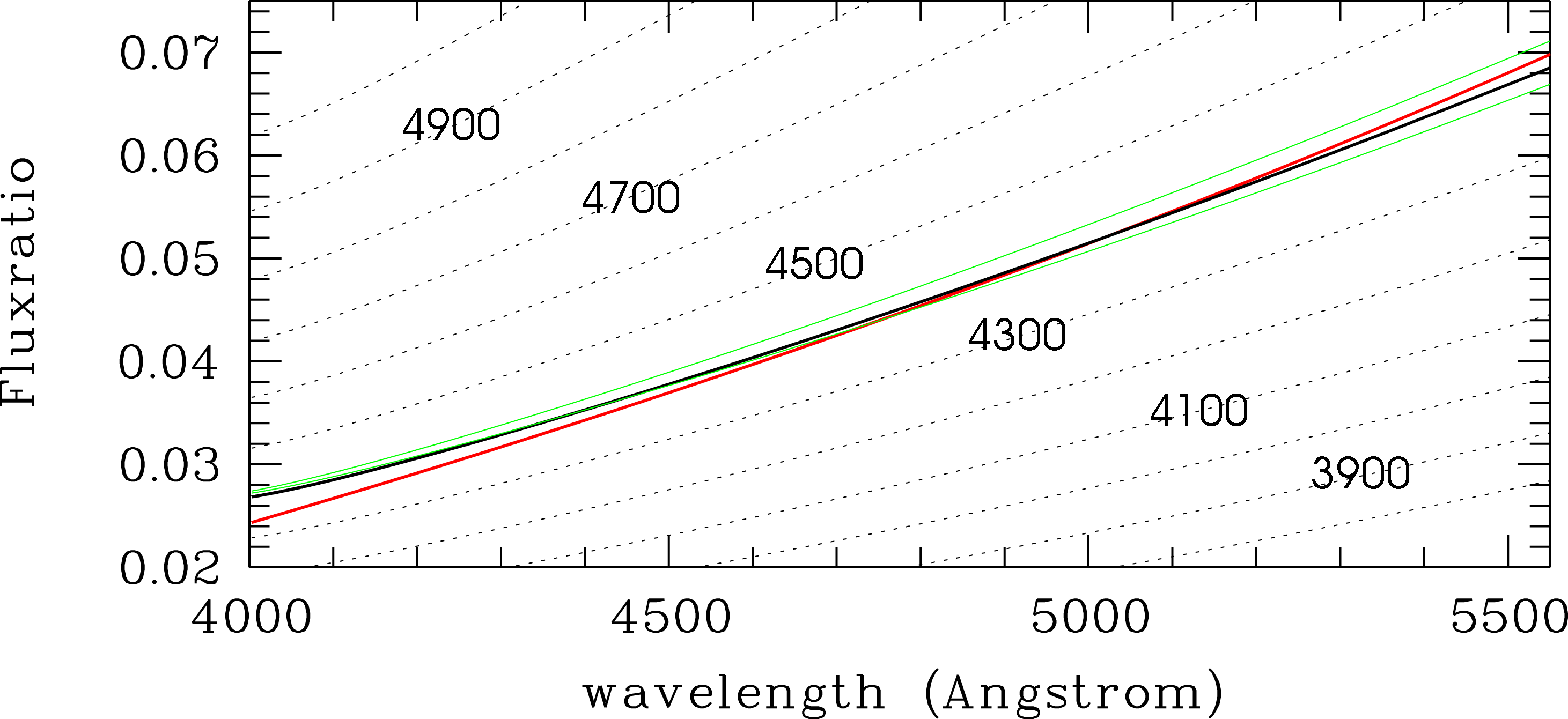}
 \caption{Continuum flux ratios between the components vs. wavelength. The dotted lines are calculated from synthetic continuum spectra with \te$_1$ of 8700\,K and \te$_2$ of 3800\,K to 5000\,K. The best agreement with the flux ratio from photometry (red line) is obtained for \te$_2$\,=\,4400\,K (black solid line). The upper green line represents \te$_1$\,=\,9000\,K, \te$_2$\,=\,4500\,K, the lower green line for \te$_1$\,=\,8400\,K, \te$_2$\,=\,4300\,K.}
 \label{Fig01}
\end{figure}

\begin{table*}
 \tabcolsep 4mm
 \caption{Atmospheric parameters of primary and secondary of RZ\,Cas derived with multiple methods.}\label{Tab02}
 \begin{tabular}{lllllll}
 \toprule
           & \multicolumn{2}{c}{Method a)} & \multicolumn{2}{c}{Method b)} & \multicolumn{2}{c}{Method c)}\\
\midrule
\te\ (K)   &  ~8650$\pm$60                 & ~4860$_{-150}^{+190}$    & ~8643$\pm$57              & ~4474$\pm$83              & ~8635$\pm$49     & ~4800$_{-120}^{+130}$         \vspace{1mm}\\
\lg\ (dex) & ~~4.41$\pm$0.09           & ~~3.7 fix                        & ~~4.42$\pm$0.07                & ~~3.7 fix                    & ~~4.41$\pm$0.06  & ~~3.7 fix                        \vspace{1mm}\\
\vt\ (\kms)& ~~3.60$_{-0.15}^{+0.32}$  & ~~1.42$\pm$0.80              & ~~~3.61$_{-0.13}^{+0.17}$ & ~~1.03$_{-0.96}^{+0.63}$ & ~~3.59$\pm$0.13  & ~~1.83$\pm$0.75                 \vspace{1mm}\\  
$[$Fe/H$]$ & $-$0.43$_{-0.06}^{+0.01}$ & $-$0.50$\pm$0.20             & $-$0.42$\pm$0.03              & $-$0.49$\pm$0.20      & $-$0.43$\pm$0.02 & $-$0.38$\pm$0.18              \vspace{1mm}\\
$[$C/H$]$  & $-$0.82$_{-0.20}^{+0.14}$ & $-$0.53$_{-0.36}^{+0.27}$& $-$0.80$_{-0.18}^{+0.13}$ & $-$0.63$_{-0.42}^{+0.29}$& $-$0.82$_{-0.18}^{+0.13}$ & $-$0.24$_{-0.46}^{+0.18}$\vspace{1mm}\\
\vs\ (\kms)& 65.40$\pm$0.95                & ~~~84.5$_{-7.9}^{+9.3}$  & 65.65$\pm$0.93            & ~~~92.3$_{-9}^{+12}$  & 65.74$\pm$0.88   & ~~~81.5$_{-7.6}^{+8.7}$   \vspace{1mm}\\
$F_2/F_1$              & \multicolumn{2}{c}{free}                             & \multicolumn{2}{c}{free}                          & \multicolumn{2}{c}{taken from photometry}                \\
$(R_2/R_1)^{\rm calc}$ & \multicolumn{2}{c}{0.926$\pm$0.008}      & \multicolumn{2}{c}{~~~~~~~1.17 fixed}    & \multicolumn{2}{c}{0.870}                           \\
$(R_2/R_1)^{\rm adj}$  & \multicolumn{2}{c}{0.780}                & \multicolumn{2}{c}{1.065}                            & \multicolumn{2}{c}{0.870}                             \\
rms (line depth)       & \multicolumn{2}{c}{0.004499}             & \multicolumn{2}{c}{0.004696}                         & \multicolumn{2}{c}{0.004525}                  \\ 
 \bottomrule
 \end{tabular}
\end{table*}

\begin{figure*}[hbt!]
 \includegraphics[width=\linewidth]{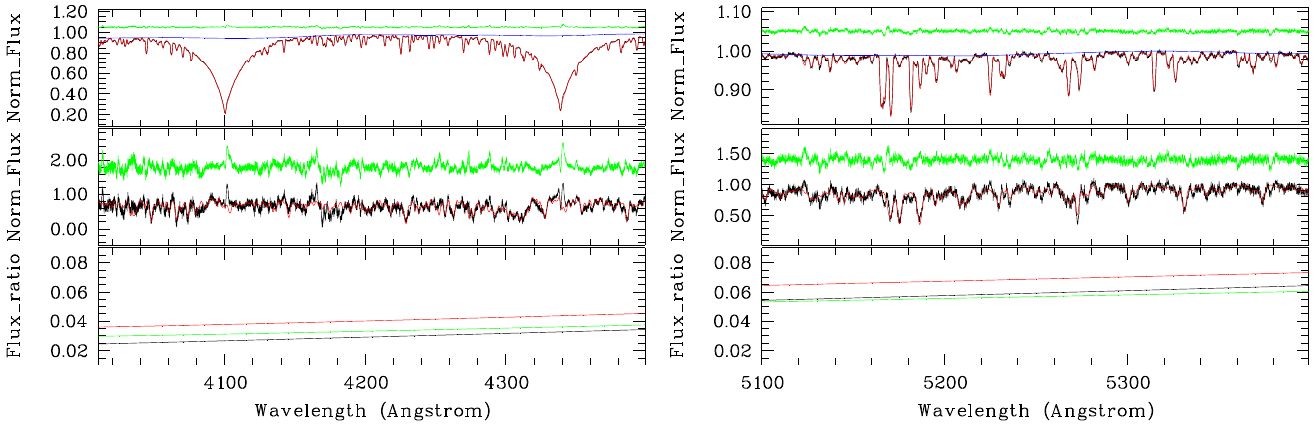}
 \caption{Results of spectrum analysis using method a) (cf. Sect.\,\ref{Sect3.3}, Table\,\ref{Tab02}), shown for the H$_{\delta}$-H$_{\gamma}$ region (left) and the region around the Mg\,I\,b triplet (right). First row: Observed, continuum adjusted composite spectrum (black), best-fitting synthetic spectrum (red), and shifted difference spectrum (green). The inverse of the applied continuum correction is shown in blue. Second row: The same for the secondary. Third row: Flux ratio of the secondary to primary from photometry (black) from spectrum analysis based on $R_2/R_1$=0.926 (red) and assuming the adjusted ratio of 0.78 (green).}
 \label{Fig02}
 \vspace{4mm}
 \includegraphics[width=\linewidth]{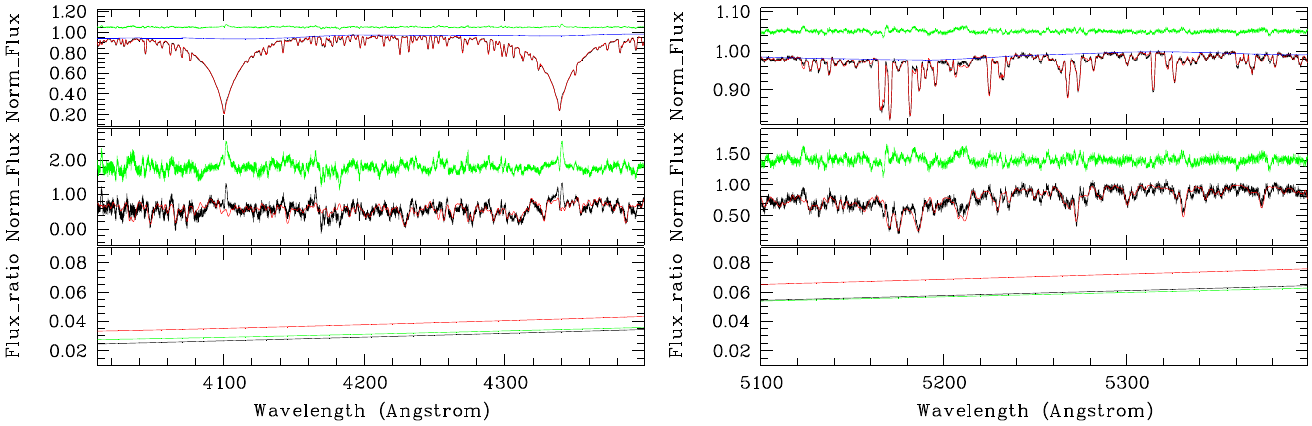}
 \caption{As Fig.\,\ref{Fig02}, but using method b). Third row: Flux ratio from photometry (black) and from spectrum analysis based on $R_2/R_1$=1.17 (red) and on the calculated ratio of 1.065 (green).}
 \label{Fig03}
 \vspace{4mm}
 \includegraphics[width=\linewidth]{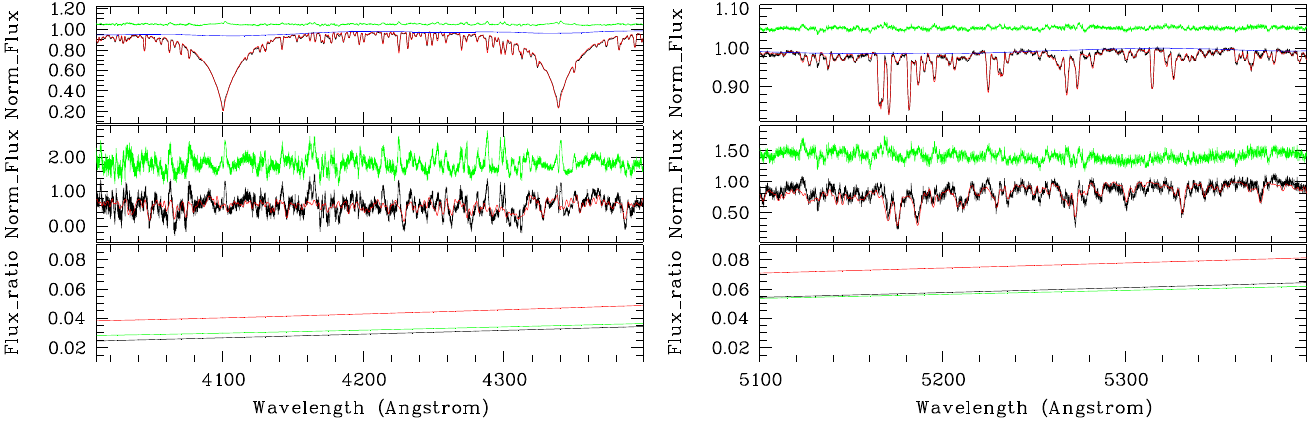}
 \caption{As Fig.\,\ref{Fig02} but using method c). Third row: Flux ratio from photometry (black) from spectrum analysis based on the photometric flux ratio assuming $R_2/R_1=1.17$ (red) and on the adjusted ratio of $R_2/R_1=0.87$ (green). Bottom row: Ratio of the photometric flux ratio to the flux ratio obtained from spectrum analysis based on $R_2/R_1=1.17$.}
 \label{Fig04}
\end{figure*}

We used the GSSP program \citep{2015A&A...581A.129T} to derive the atmospheric parameters of the components of RZ\,Cas. It is based on the spectrum synthesis method and performs a grid search in stellar parameters. Synthetic spectra were computed with SynthV \citep{1996ASPC..108..198T}, based on a library of atmosphere models computed with LLmodels \citep{2004A&A...428..993S} for the hot primary and on MARCS models \citep{2008A&A...486..951G} for the cool secondary. Atomic data were taken from the VALD database \citep{2000BaltA...9..590K}.

Besides the atmospheric parameters, GSSP also has to solve for the a priori unknown flux ratio of the components. In the following, we use the continuum flux ratio of the secondary (component\,2) to  primary (component\,1) for comparison.  It is interpolated from the $uvby$ luminosities provided by \citet{2004MNRAS.347.1317R} and shown by the red line in Fig.\,\ref{Fig01}. To get an impression of the influence of the \te\ of primary and secondary on the resulting flux ratios with wavelength, we computed synthetic continuum spectra for different \te\ of the components, assuming \lg$_2$\,=\,3.7  \citep{2004MNRAS.347.1317R}, and \lg$_1$\,=\,4.4 and [Fe/H]=$-$0.42 as derived in Sect.\,\ref{Sect3.3}. Synthetic flux ratios were computed from the ratio of the continuum spectra assuming a radii ratio of $R_2/R_1$\,=\,1.17, also taken from \cite{2004MNRAS.347.1317R}. The best representative theoretical curve is shown by the black line in Fig.\,\ref{Fig01} and corresponds to $T_{\rm eff1}$\,=\,8700\,K, $T_{\rm eff2}$\,=\,4400\,K. From the green lines, we see that a variation of \te$_2$ by 100\,K requires a variation of \te$_1$ by 300\,K to approximately fit the photometric flux ratio.

We applied three different methods (labelled a, b, and c in what follows) to determine the atmospheric parameters together with the radii ratio and continuum flux ratio of the secondary to primary.

a) We use the standard method of the GSSP program that takes the wavelength dependence of the flux ratio from the ratio of the synthetic continuum spectra calculated with SynthV and scales it with the square of the radii ratio. The latter is obtained from comparing the observed with the synthetic normalised spectra on a grid of atmospheric parameters. To obtain the atmospheric parameters together with the radii ratio, we minimise
\begin{equation}
 \chi^2 = \sum_\lambda\left(\frac{\d o(\lambda)-s(\lambda)}{\d\sigma}\right)^2,
 \label{GSSP1}
\end{equation}
 where $o(\lambda)$ and $s(\lambda)$ are the observed and synthetic composite spectra, respectively, and $\sigma$ is the estimated mean error of $o(\lambda)$. The synthetic spectrum is computed from
\begin{equation}
 s(\lambda) = \frac{\d s_1(\lambda,v_1)+V_F(\lambda)\,s_2(\lambda,v_2)}{\d 1+V_F(\lambda)},
 \label{GSSP2}
\end{equation}
where $s_1$ and $s_2$ are the synthetic spectra of the components shifted for their RVs $v_1, v_2$ (all spectra in line depths), and the flux ratio
\begin{equation}
 V_F(\lambda) = \left(\frac{\d R_2}{\d R_1}\right)^2\frac{\d C_2(\lambda)}{\d C_1(\lambda)}\label{GSSP3}    
\end{equation}
 is the product of the squared radii ratio and the ratio of the continuum fluxes $C_2$ and $C_1$ per unit surface, determined with SynthV.

b) We use Equations\,\ref{GSSP1} to \ref{GSSP3} as before, but fix the radii ratio to 1.17 as obtained from the uvby photometry.

c) We replace the flux ratio $V_F$ computed so far from Eq.\,\ref{GSSP3} by the flux ratio obtained from the $uvby$ photometry. No synthetic spectra are needed in this case. The best-fitting radii ratio can then be determined from Eq.\,\ref{GSSP3} using a simple least-squares fit.

Because of the known degeneracy between various free parameters, we reduced their number. The spectrum of the secondary with its small contribution to total light suffers from low S/N and we fixed its \lg\ to 3.7, as derived from light curve analysis \citep{2004MNRAS.347.1317R}. Furthermore, we used the same values for its elemental abundances as derived for the primary, except for the iron and carbon abundances, which we determined separately. For the primary, we basically know the photometric value of \lg\,=\,4.33(3) from \citet{2004MNRAS.347.1317R}. Because of the good S/N of the spectra and the fact that the Balmer lines shapes are very sensitive to \lg\ in this temperature range, we decided to use \lg\ of the primary as a free parameter to compare the results with the photometric results. 

The GSSP allows us to iterate atmospheric parameters such as \te, \lg, and microturbulent velocity \vt\ together with \vs\ and the surface abundance of one chemical element at the same time. We started, in this way, to optimise [Fe/H] together with the other parameters, first for the primary, then for the secondary, and repeated until the change in parameters and $\chi^2$ became marginal. In a next step, we fixed all parameters to the values obtained so far and optimised the elemental abundances of all other chemical elements for which significant contributions in the spectra could be found.

\subsection{Results}\label{Sect3.3}

Tables\,\ref{Tab02} and \ref{Tab03} list the results. $(R_2/R_1)^{\rm calc}$ in Table\,\ref{Tab02} is the radii ratio obtained from the analysis using methods a, b), or c). $(R_2/R_1)^{\rm adj}$ is the adjusted radii ratio that follows when we apply the least-squares fit based on Eq.\,\ref{GSSP3}, as used in method c) for the two other methods as well. This means that when keeping all other parameters obtained with the various methods, the radii ratio has to be changed from $(R_2/R_1)^{\rm calc}$ to $(R_2/R_1)^{\rm adj}$  to give the best agreement with the photometric flux ratio. For method c), both radii ratios are identical. The quantity $(R_2/R_1)^{\rm adj}$ is used in the following as a measure for the agreement of the spectroscopic with the photometric results. It should be equal to 1.17 in the optimum case.

Figures\,\ref{Fig02} to  \ref{Fig04} compare the results from the three methods. These figures show in the first row the observed spectrum together with the best-fitting synthetic composite spectrum. The second row compares the ``observed'' spectrum of the secondary with the best-fitting synthetic spectrum found for the secondary. The observed spectrum of the secondary was therefore computed by subtracting the best-fitting spectrum of the primary from the observed composite spectrum. These spectra are rescaled according to the obtained flux ratio and the observed spectrum of the secondary is correspondingly noisy. The bottom row compares the obtained flux ratio as a function of wavelength with that obtained from photometry. The differences seen by eye are marginal. 

From Table\,\ref{Tab02} and Figures\,\ref{Fig02} to \ref{Fig04} we conclude as follows:
First, there are only small differences in the atmospheric parameters and elemental abundances derived with the different methods for the primary.
Second, methods a) and c) give \te\ of the secondary that agree with each other within the 1\,$\sigma$ error bars, but both are distinctly higher than the value expected from photometry.  The \te\ derived from method b), on the other hand, is in agreement with that value.
Third, the iron and carbon abundances of the secondary cannot be distinguished from those of the primary within the 1$\sigma$ errors (Table\,\ref{Tab02}), but all three methods yield a distinctly lower [C/H] of the primary component compared to the solar value. 
Fourth, all three methods deliver flux ratios higher than the photometric ratio, as can be seen from the bottom panels in Figures\,\ref{Fig02} to \ref{Fig04}.  Based on the atmospheric parameters derived with methods a) and c), RZ\,Cas should have much smaller radii ratios (the $(R_2/R_1)^{\rm adj}$ in Table\,\ref{Tab02}) than computed. From the results of light curve analysis by \citet{2004MNRAS.347.1317R}, however, we expect a radii ratio of 1.17 and can exclude that the radius of the secondary is larger than that of the primary. From method b), on the other hand, we obtain a value of $(R_2/R_1)^{\rm adj}$ that is larger than unity and not so far from 1.17.

\begin{table}
 \tabcolsep 3.3mm
 \caption{Elemental abundances. We list the solar values based on  \citet{2009ARA&A..47..481A}, given as 12\,+\,log{(E/H)}, and the abundances of the primary of RZ\,Cas relative to the solar values, measured with the multiple methods.}\label{Tab03}
\begin{tabular}{lcccccc}
\toprule
 & Solar & Method a) & Method b) & Method c) \\
\midrule
C   &8.43 & $-0.82_{-0.20}^{+0.14}$ & $-0.80_{-0.18}^{+0.13}$ & $-0.82_{-0.18}^{+0.13}$\vspace{1mm}\\
O   &8.69 & $-0.11_{-0.34}^{+0.21}$ & $-0.22_{-0.48}^{+0.24}$ & $-0.17_{-0.39}^{+0.22}$\vspace{1mm}\\
Mg  &7.60 & $-0.14_{-0.05}^{+0.05}$ & $-0.17_{-0.05}^{+0.05}$ & $-0.17_{-0.05}^{+0.05}$\vspace{1mm}\\
Si  &7.51 & $-0.20_{-0.14}^{+0.12}$ & $-0.18_{-0.14}^{+0.13}$ & $-0.19_{-0.14}^{+0.13}$\vspace{1mm}\\
Ca  &6.34 & $-0.39_{-0.08}^{+0.07}$ & $-0.38_{-0.08}^{+0.08}$ & $-0.39_{-0.07}^{+0.07}$\vspace{1mm}\\
Sc  &3.15 & $-0.23_{-0.09}^{+0.09}$ & $-0.21_{-0.09}^{+0.09}$ & $-0.24_{-0.09}^{+0.09}$\vspace{1mm}\\
Ti  &4.95 & $-0.18_{-0.04}^{+0.04}$ & $-0.20_{-0.04}^{+0.04}$ & $-0.21_{-0.04}^{+0.04}$\vspace{1mm}\\
V   &3.93 & $-0.06_{-0.20}^{+0.15}$ & $-0.11_{-0.22}^{+0.16}$ & $-0.07_{-0.19}^{+0.15}$\vspace{1mm}\\
Cr  &5.64 & $-0.35_{-0.07}^{+0.06}$ & $-0.36_{-0.06}^{+0.06}$ & $-0.36_{-0.06}^{+0.06}$\vspace{1mm}\\
Mn  &5.43 & $-0.45_{-0.13}^{+0.11}$ & $-0.47_{-0.14}^{+0.12}$ & $-0.46_{-0.13}^{+0.11}$\vspace{1mm}\\
Fe  &7.50 & $-0.43_{-0.06}^{+0.01}$ & $-0.42_{-0.04}^{+0.02}$ & $-0.43_{-0.02}^{+0.03}$\vspace{1mm}\\
Ni  &6.22 & $-0.41_{-0.15}^{+0.12}$ & $-0.41_{-0.15}^{+0.12}$ & $-0.43_{-0.15}^{+0.12}$\vspace{1mm}\\
Y   &2.21 & $-0.32_{-0.26}^{+0.19}$ & $-0.32_{-0.26}^{+0.20}$ & $-0.32_{-0.26}^{+0.19}$\vspace{1mm}\\
Ba  &2.18 & $-0.45_{-0.18}^{+0.17}$ & $-0.42_{-0.18}^{+0.17}$ & 
$-0.45_{-0.17}^{+0.17}$\\            
 \bottomrule                    
 \end{tabular}                  
\end{table}

There is a large degeneracy between various parameters such as \te, \lg, \vt, [Fe/H] (both components), [Mg/H] (cool component), and radii ratio. This explains why we end up in both methods a) and c) with extraordinary small radii ratios on costs of higher \te\  of the secondary (much too high when comparing with Fig.\,\ref{Fig01}). We assume that the number of degrees of freedom is too high when trying to optimise the radii ratio together with the other parameters; the signal from the faint companion in our composite spectra is simply too low for that. Thus we observe, although it does not give the smallest root mean square (rms) of the O-C residuals, that the most reliable results are obtained when fixing the flux ratio to the photometric ratio as we did in method b). 

\section{Radial velocities and LSD profiles with LSDbinary}\label{Sect4}

\begin{figure}\centering
 \includegraphics[width=\linewidth]{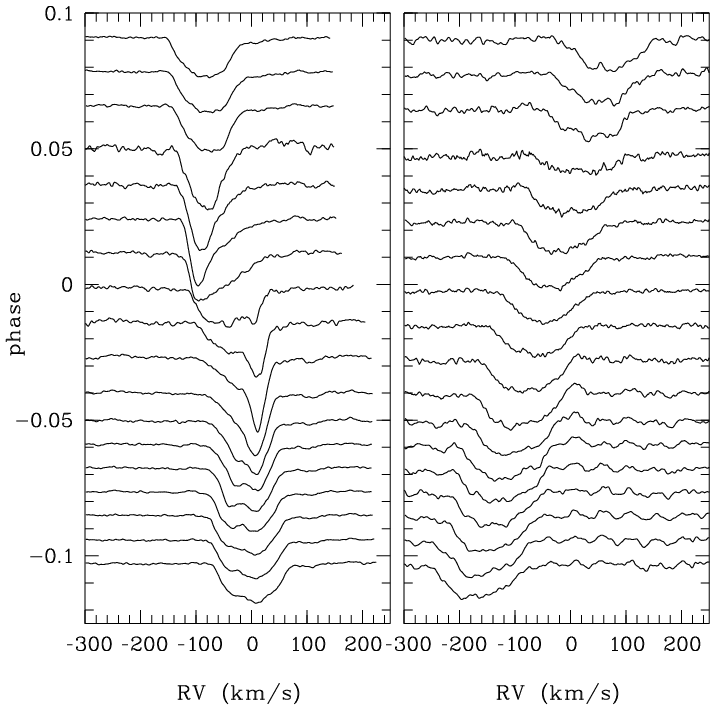}
 \caption{LSD profiles computed with LSDbinary from spectra taken during primary eclipse for the primary (left) and secondary (right) of RZ\,Cas. Phase zero corresponds to Min\,I.}
 \label{Fig05}
\end{figure}

The classical method of LSD \citep[][]{1997MNRAS.291..658D} is based on using one line mask as a template. In the result,  a strong deconvolved line profile is obtained for the star that best matches this template, whereas the contribution of the other component is more or less suppressed. \citet{2013A&A...560A..37T} generalised the method so that it  can simultaneously compute an arbitrary number of LSD profiles from an arbitrary number of line masks. In this work, we used the LSDbinary program written by V. Tsymbal, which computes separated LSD profiles for the two stellar components using as templates two synthetic spectra that are based on two different atmosphere models. The program also delivers optimised values of the RVs of the components and their radii ratio. A first successful application of the program to the short-period Algol \object{R CMa} is presented in \cite{2018A&A...615A.131L}; this work also includes a short description of the program algorithms and shows the advantages of LSDbinary against TODCOR \citep{1994ApJ...420..806Z} in the case of very small flux ratios between the components of binary stars.

We applied LSDbinary to the spectra of RZ\,Cas to obtain the separated LSD profiles and RVs of its components. Figure\,\ref{Fig05} shows LSD profiles computed from RZ\,Cas spectra taken during the primary eclipse in 2016 as an example. The RME can clearly be seen in the profiles of the primary, as well as the RV variation due to orbital motion in the profiles of both components. We will use the obtained RVs and LSD profiles for a detailed investigation of the stellar and system parameters of RZ\,Cas in this first part and of the pulsations of its primary component in Paper\,II.

\section{Rotation velocity of the primary}\label{Sect5}

We applied three different methods to determine  \vs\ of the primary of RZ\,Cas. In all cases, we only use spectra taken in out-of-eclipse phases.

\subsection{Fourier method}\label{Sect5.1}
First, we applied the Fourier method \citep{1933MNRAS..93..478C, 1976PASP...88..809S, 2005oasp.book.....G} to the calculated LSD profiles. This method (DFT hereafter) is based on the determination of the rotation-broadening related zero points in the Fourier power spectra of the profiles. We assumed a linear limb darkening law with a limb darkening coefficient $\beta$\,=\,1.5 (or $b$\,=\,0.6 with $b=\beta/(1+\beta)$). Varying the value of $\beta$ leads to slight systematic effects in \vs, but changes were minor compared to the differences to the results from the two other methods described below. The selection of zero points was based on a 3$\sigma$ clipping, comparing the \vs\ from single zero points with the mean values from all zero points.

\begin{figure}\centering
 \includegraphics[angle=-90, width=\linewidth]{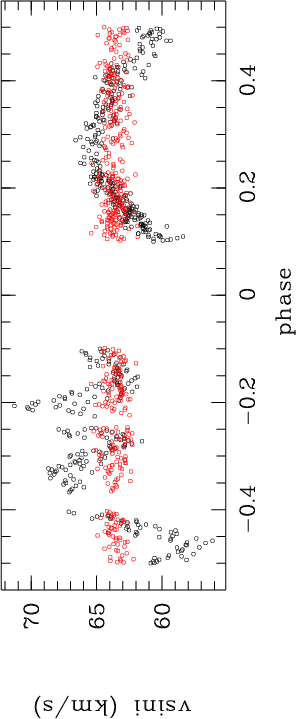}
 \caption{Values of \vs\ obtained from DFT vs. out-of-eclipse phases measured from the LSD profiles observed in 2016 (black) and from the same profiles corrected for the LPV found with FAMIAS (red).}
 \label{Fig06}
\end{figure}

\noindent In the result, we obtained \vs\ values strongly varying with orbital phase. Figure\,\ref{Fig06} shows an example.  We assume that this variation comes from Algol-typical effects such as an inhomogeneous circum-primary gas-density distribution and/or surface structures caused by the influence of mass transfer and gas stream from the secondary. In a next step, we removed all the variations found in the line profiles with the pixel-by-pixel method of FAMIAS \citep{2008CoAst.157..387Z} by subtracting all frequency contributions computed with the mentioned program from the profiles. This method will be explained in detail in Paper\,II. For each season, we considered a certain pixel of all LSD profiles as part of one time series from which we subtracted the found contributions and did this pixel-by-pixel to build the undistorted profiles. We performed that for all seasons but 2008 for which we did not have enough data to apply the pixel-by-pixel method. The subtraction of the high-amplitude, low-frequency contributions found with FAMIAS (most of them are harmonics of the orbital frequency) are responsible for the cleaning, not the faint high-frequency oscillations due to pulsation.  The resulting \vs\ are shown in Fig.\,\ref{Fig06} in red. The distribution is now much flatter and we used it to calculate the \vs\ of the corresponding season as its arithmetic mean. Results are listed in the third column of Table\,\ref{Tab04}.

\subsection{Single spectral line} 
\begin{figure}\centering
 \includegraphics[width=.82\linewidth]{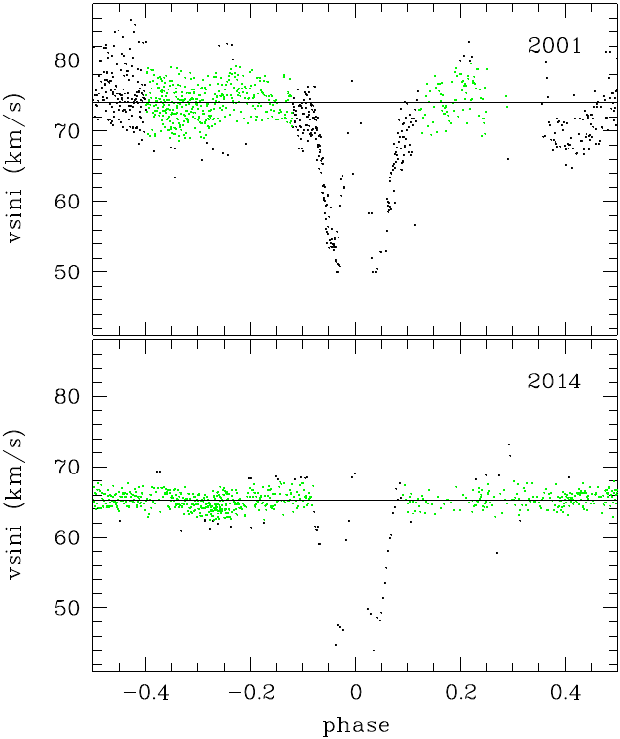}
 \caption{Values of \vs\ determined from the Fe\,II 5317\,\AA\ line vs. orbital phase, shown for 2001 and 2014. The mean values were built from the values indicated in green.}
 \label{Fig07}
\end{figure}

Next, we looked for stronger spectral lines of the primary in the composite spectra that are free of blends from the cool companion. We found only one line consisting of the Fe\,II doublet 5316.61/5316.78\,\AA\ that fulfils that condition. For each season, we fitted the line profiles by synthetic spectra computed with the parameters listed in Table\,\ref{Tab02}, method b) to determine the best-fitting \vs\ and its error. This means that we were fixing all atmospheric parameters except for  \vs\ to the solution determined from the spectra observed in 2009 when the star was in a relatively quiet phase. However, this approach does not account for the effects in active phases of RZ\,Cas such as the attenuation of light by circumstellar material as found for the year 2001 for example \citep{2009A&A...504..991T}. Thus we used two more free parameters in our fit, correction factor $a$ counting for different line depths caused by the presumed effects, and factor $b$ to correct the continuum of the observed spectrum in the vicinity of the Fe\,II line.  Both were determined from a least-squares fit
\begin{equation}
    \{1-[1-a\,P_s(v\sin{i})]-b\,P_o\}^2~\longrightarrow~min.
\end{equation}
where $P_s$ is the synthetic and $P_o$ the observed line profile. Finally, we built the mean \vs\ per season from the weighted mean of well-selected data points using  3$\sigma$ clipping. Figure\,\ref{Fig07} shows two examples: one for 2001 in which the star was in an active phase, and one for 2014 in which the \vs\ shows a much smoother behaviour with orbital phase. The results are listed in the fourth column of Table\,\ref{Tab04}.

\subsection{Using FAMIAS}
In Paper\,II, we will 
use the moment methods \citep{1992A&A...266..294A} of the FAMIAS program for a mode identification of low-$l$ degree modes. For that, we cleaned the observed LSD profiles for all low-frequency distributions in the same way as described in Sect.\,\ref{Sect5.1}. One free parameter in applying the moment method is the \vs\ of the primary, which optimum value and 1$\sigma$ error we obtained from the resulting $\chi^2$-distribution. The values are listed in the last column of Table\,\ref{Tab04}. The number of observations in 2008 were not sufficient to apply this method.

\subsection{Comparison}\label{Sect5.4}
\begin{figure}\centering
\includegraphics[angle=-90, width=\linewidth]{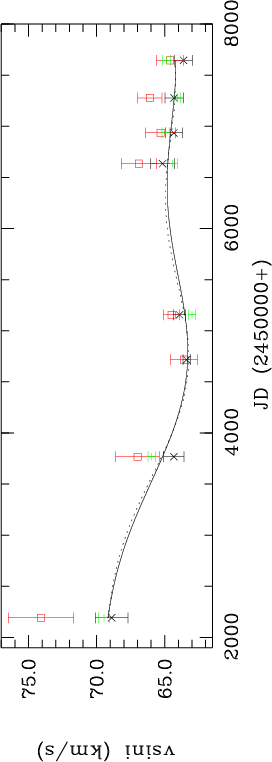}
\caption{Values of \vs\  measured from a) DFT (black crosses), b) the Fe\,II line (red squares), and c) FAMIAS (green plus signs). The solid and dotted lines show possible sinusoidal variations (see text).}
\label{Fig08}
\end{figure}

The \vs\ determined with the three different methods are shown in Fig.\,\ref{Fig08}. The values obtained from DFT and FAMIAS agree in most cases within the 1$\sigma$ error bars; there is a larger difference for 2006. A systematic offset can be observed for the \vs\ values obtained from the Fe\,II line. This is in particular the case for 2001 when RZ\,Cas was in an active phase. We assume that the offset comes from the cleaning procedures that we applied when using the other two methods. In these cases we removed the low-frequency contributions that we found with the pixel-by-pixel method of FAMIAS from the LSD profiles so that we removed the line broadening effects due to orbital-phase dependent dilution effects by circumstellar material in this way. 

\begin{table}
\tabcolsep 1.4mm
\caption{Values of \vs\  obtained from the three methods in multiple years. JD is JD\,2\,450\,000+. The last column gives the rotation-to-orbit synchronisation factor. Here and in the following, values in parentheses are the errors in units of the last digits.}\label{Tab04}
\begin{tabular}{cccccc}
\toprule
Year &  JD & \vs$_{\rm DFT}$ &\vs$_{\rm FeII}$ & \vs$_{\rm FAMIAS}$ & $F_1$ \\
     &   &  (\kms)        & (\kms)          & (\kms) & \\
\midrule
2001 & 2192 & 68.9(1.2) & 74.1(2.4) & 69.47(39) & 0.978(15)\\
2006 & 3768 & 64.33(74) & 67.0(1.6) & 65.95(28) & 0.921(12)\\
2008 & 4717 & 63.39(27) & 63.6(1.0) &  --       & 0.896(11)\\
2009 & 5156 & 63.91(45) & 64.49(59) & 62.99(25) & 0.897(11)\\
2013 & 6635 & 65.16(87) & 66.9(1.3) & 64.43(35) & 0.916(13)\\
2014 & 6936 & 64.33(61) & 65.3(1.1) & 64.94(26) & 0.914(12)\\
2015 & 7278 & 64.31(67) & 66.11(91) & 64.08(23) & 0.908(12)\\
2016 & 7646 & 63.62(65) & 64.6(1.0) & 64.87(29) & 0.909(12)\\
\bottomrule
\end{tabular}
\end{table}

Finally, we built a spline fit through the averaged values obtained from DFT and  FAMIAS and interpret this as the typical variation of \vs\ over the epoch of our observations.
We see that \vs\ was distinctly larger in 2001 compared to the other seasons. We assume that its value increased during the active phase of RZ\,Cas near to 2001 because of the acceleration of the outer layers of  the primary by angular momentum transport via mass transfer from the cool component (see Sect.\,\ref{Sect8} for a more detailed discussion). A period search in the values averaged from DFT and FAMIAS gives a period of 9.6\,yr, overlaid on a long-term trend (the solid line in Fig.\,\ref{Fig08}). A fit based on the 9.0\,yr period as discussed in Sect.\,\ref{Sect7.2.3} is shown by the dotted line. The difference is marginal. The last column in Table\,\ref{Tab04} lists the rotation-to-orbit synchronisation factor of the primary, $F_1$, calculated from its radius, taken from \citet{2004MNRAS.347.1317R} as 1.67\,$\pm$\,0.02\,R$_\odot$, and the arithmetic mean of the \vs\ measured with DFT and FAMIAS. The result shows that the primary rotates sub-synchronously, only in 2001 it reaches almost synchronous rotation velocity.

\section{Orbital period changes}\label{Sect6}

It is hard to search for orbital period changes in the range of a few seconds when the orbital RV curves are heavily distorted by non-Keplerian effects such as in the case of RZ\,Cas  (see Fig.\,\ref{Fig13}). Table\,\ref{Tab05} lists the orbital periods and times of minimum ($T_{\rm Min}$ hereafter) derived from the RVs in different seasons using the method of differential corrections \citep{1910PAllO...1...33S, 1941PNAS...27..175S}. The 1$\sigma$ errors of the period are of the order of two seconds or more, which is larger than the expected period changes.  Also the inclusion of effects such as Roche geometry of the components and spots on the stellar surfaces into the PHOEBE calculations (Sect.\,\ref{Sect7.2}) did not help us reach the desired accuracy in orbital period. The $T_{\rm Min}$, on the other hand, could be very precisely determined.

\begin{table}\centering
\tabcolsep 1.3mm
\caption{Orbital periods $P_{RV}$ and times of minimum $T_{\rm RV}$ derived from the RVs from single seasons, and period changes $\triangle P_{phot}$ and periods $P_{phot}$ derived from the photometric $T_{\rm Min}$.}
\label{Tab05}
\begin{tabular}{cllrc}
\toprule
Year  & \multicolumn{1}{c}{$P_{RV}$}                  & \multicolumn{1}{c}{$T_{\rm RV}$} 
      & \multicolumn{1}{c}{$\triangle P_{phot}$} & \multicolumn{1}{c}{$P_{phot}$}\\
      & \multicolumn{1}{c}{(d)}                    & \multicolumn{1}{c}{BJD\,2450000+}
      & (sec)                                      & (d) \\
\midrule
2001 &  1.19572(30)  & 2190.9954063(32) &    1.07 & 1.1952626\\
2006 &  1.195248(20) & 3775.9076591(23) & $-$0.15 & 1.1952486\\
2008 &  1.19525(26)  & 4717.7661978(51) & $-$0.03 & 1.1952500\\
2009 &  1.1954(13)   & 5156.4217537(36) &    0.40 & 1.1952549\\
2013 &  1.195307(65) & 6663.6333727(22) &    0.31 & 1.1952539\\
2014 &  1.19533(28)  & 6936.1537115(17) &    0.45 & 1.1952554\\
2015 &  1.195278(17) & 7279.1944163(12) &    0.70 & 1.1952583\\
2016 &  1.19497(26)  & 7644.9432256(20) &    0.67 & 1.1952580\\
\bottomrule
\end{tabular}
\end{table}

\begin{figure}
\includegraphics[angle=-90, width=.9\linewidth]{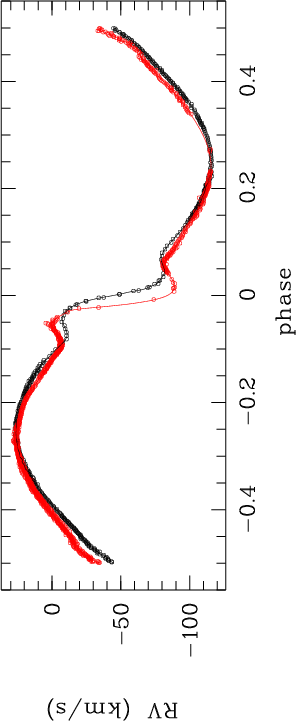} 
\caption{Radial velocities of the primary in 2006 (black) and 2001 (red) vs. orbital phase. Phase is calculated from $P$ and $T_{\rm min}$ as derived from the RVs in 2006.}
\label{Fig09}
\end{figure}

\begin{figure}
\includegraphics[angle=-90, width=\linewidth]{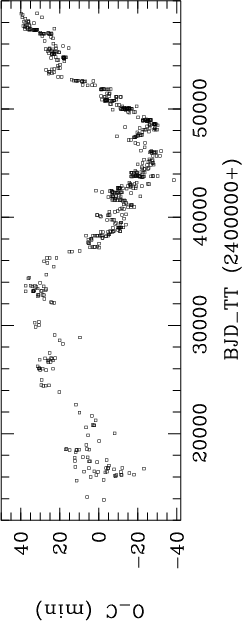}
\caption{O-C values calculated from the corrected $T_{\rm Min}$, full range in JD.}
\label{Fig10}
\end{figure}

When we plot the RVs versus orbital phase where the latter is calculated from $T_{\rm min}$ and $P$ of one single season, we see a clear phase shift in RV compared to other seasons (Fig.\,\ref{Fig09}). Since we do not have any information about the behaviour of the system in between, we cannot deduce period shifts from our RVs alone, however. To fix the problem, we used the photometric $T_{\rm Min}$ from literature. We collected data from the O-C Gateway\footnote{http://var2.astro.cz/ocgate/index.php?lang=en}, Bob Nelson's Data Base of O-C Values\footnote{w.aavso.org/bob-nelsons-o-c-files}, $T_{\rm Min}$ kindly provided by J.\,Kreiner\footnote{https://www.as.up.krakow.pl/o-c/} \citep[also see][]{2004AcA....54..207K}, and unpublished data from D.\,Mkrtichian. Cross-checking for duplicates and rejecting all visual observations, we ended up with 605 $T_{\rm Min}$ covering the time span from 1896 to 2019, to which we added the eight $T_{\rm Min}$ derived from our spectra. We converted all dates given as HJD to BJD based on terrestial time TT. Then we computed the overall best fitting period from a least-squares fit $T_{\rm Min}$ versus season number $E$, yielding $P_0$\,=\,1.195250392(61)\,d and $T_0$\,=\,2\,453\,775.89453(79). The resulting values
 \begin{equation}
     O-C = T_E-T_0-P_0E
 \end{equation}
are shown in Fig.\,\ref{Fig10}. 

There exist different approaches to determine the local period from an O-C diagram. The classic method is to fit segments by linear or parabolic functions to calculate constant periods or linearly changing periods per segment, respectively. But there are also approaches that assume a continuous change of the orbital period like that of \citet{1994A&A...282..775K}, see \citet{2001OAP....14...91R} for an overview on other methods. We applied a similar procedure as used in \citet{2018MNRAS.475.4745M}, interpolating the O-C data to a grid of step width one in season $E$, using spline fits together with 3$\sigma$-clipping and computing the period change from the local slope of the resulting fit. It is
 \begin{equation}\label{deltaP}
     (O-C)_E-(O-C)_{E-1} = T_E-T_{E-1}-P_0
 \end{equation}
where $T_E$\,$-$\,$T_{E-1}$ is the local period and thus Eq.\,\ref{deltaP} describes the local difference $\triangle P$\,=\,$P$\,$-$\,$P_0$. A necessary precondition for our method is that the observed data points are sufficiently dense so that the spline fit gives a reliable prediction of the behaviour between these points. For that reason, we only considered all $T_{\rm Min}$ observed after BJD 2\,437\,000.

\begin{figure}
\includegraphics[width=\linewidth]{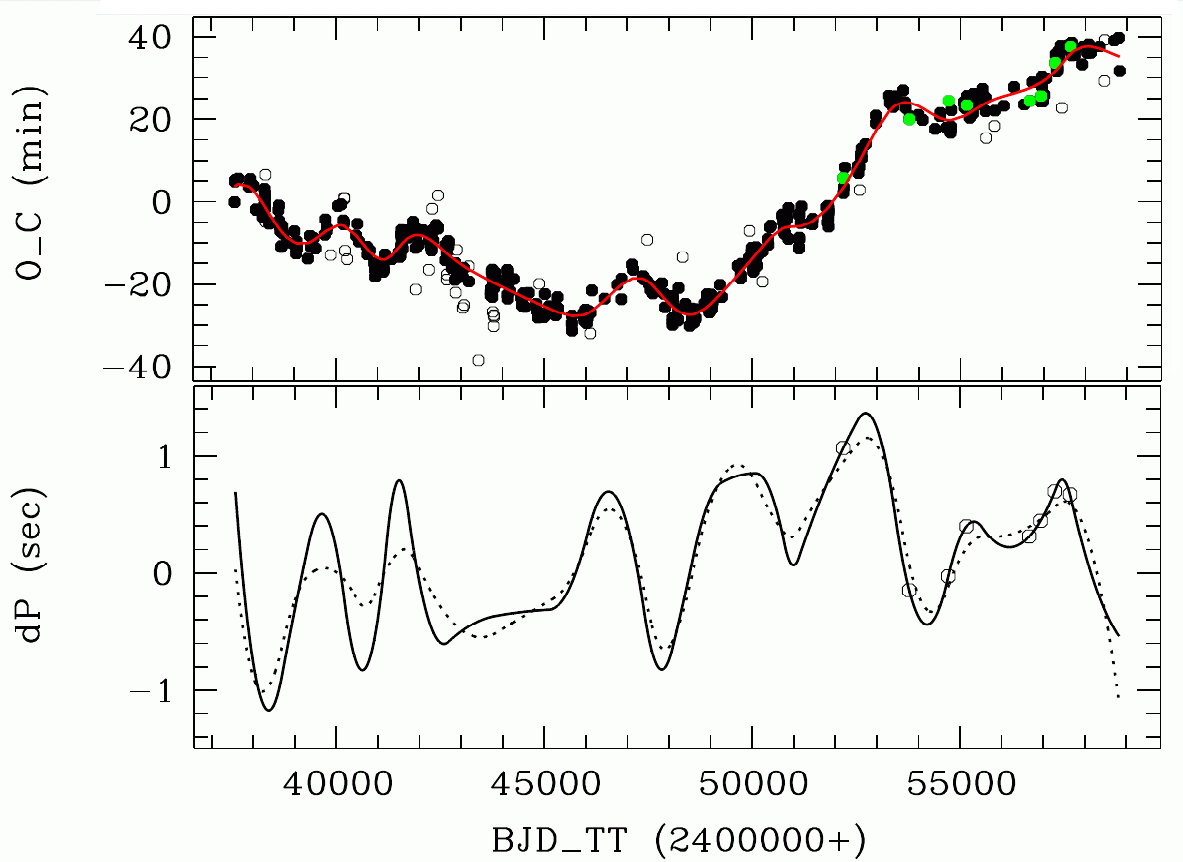}
\caption{O-C values and derived period changes. Top: O-C values (filled black circles) fitted by splines (red line). Values considered as outliers are shown by open circles and those derived from our spectra in green. Bottom: Period changes calculated from the local slope of the spline fit (see text). Open circles indicate the seasons of our spectroscopic observations.}
\label{Fig11}
\end{figure}

Figure\,\ref{Fig11} shows in its top panel the O-C values together with the spline fit. The $T_{\rm Min}$ obtained from our RVs ($T_{\rm RV}$ in Table\,\ref{Tab05}) are shown in green and fit very well. The bottom panel shows the resulting period changes by the solid line, where we assumed a ``best-selected'' smoothness parameter for the underlying spline fit. To give an impression of the influence of the smoothness parameter, we also show (dotted line) the period changes resulting from a much larger parameter, leading to a  more relaxed spline fit. The positions of our spectroscopic observations are indicated. Obtained period changes and periods are listed in Table\,\ref{Tab05} as $\triangle P_{LC}$ and $P_{LC}$. 

\begin{figure}\centering
\includegraphics[angle=-90, width=\linewidth]{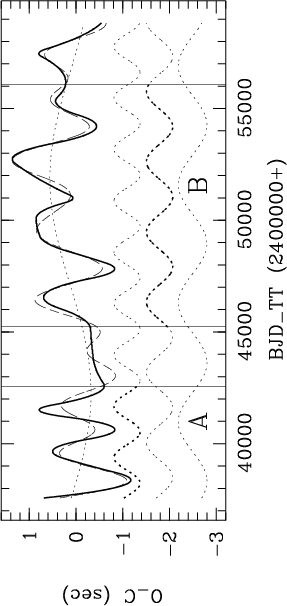}
\caption{Calculated period changes (solid line) fitted by six frequencies (dashed line). The dotted lines show (from top to bottom) the contributions of the 52.7, 6.3, 8.6, and 14.5\,yr periods.}
\label{Fig12}
\end{figure}

\begin{figure*}\centering
\includegraphics[width=0.76\textwidth]{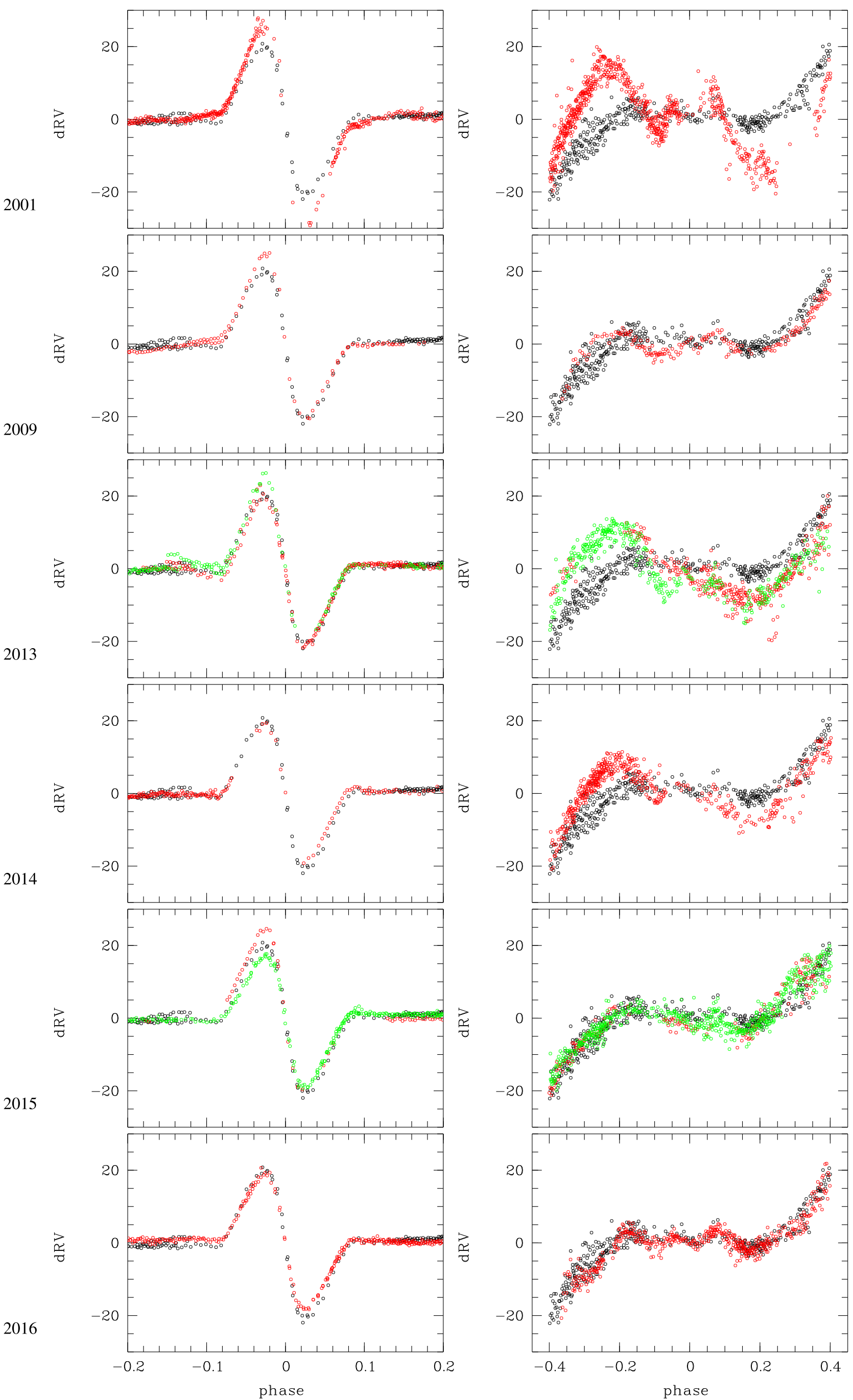}
\caption{Radial velocity residuals. Left: the Rossiter-McLaughlin effect in the RVs of the primary in different years (red) after subtracting the best-fitting Keplerian orbits. For comparison, the RVs from 2006 are shown in black. In 2013, the RVs are divided into BJD\,<\,2\,456\,600 (2013a, red) and BJD\,>\,2\,456\,600 (2013b, green), and in 2015 into BJD\,<\,2\,457\,250 (2015a, red) and BJD\,>\,2\,457\,250 (2015b, green). Right: The same for the RVs of the secondary, shown over a larger range in orbital phase. Phase zero corresponds to Min\,I in each case.}
\label{Fig13}
\end{figure*}

\citet{2018MNRAS.475.4745M} derived typical timescales of 4.8, 6.1, and 9.2\,yr from the O-C variations. We did a frequency search in the obtained period changes using the PERIOD04 program \citep{2005CoAst.146...53L} and found the six periods listed in Table\,\ref{Tab06}. We do not assume that the observed variation can be described as strictly cyclic but consider the found periods as typical timescales that describe the behaviour in certain seasons. Figure\,\ref{Fig12} illustrates this. It shows the calculated period changes together with the best fit of the six periods and also single contributions from four of the six periods. The longest period of 52.7\,yr describes a long-term trend in the data. The 6.3\,yr period describes the variations before BJD 2\,442\,560 (segment A), and the 8.6\,yr period the behaviour between 2\,445\,240 and 2\,456\,070 (segment B). The 14.5\,yr period is responsible for an amplitude variation over a longer timescale. The remaining two periods of 6.9 yr and 10.8\,yr cannot directly be linked to the variations in this way, but improve the fit by counting for their non-cyclicity. When doing a separate frequency search in segments A and B, we find the dominant periods as 5.9\,yr and 8.4\,yr, respectively. This shows that we can only estimate the order of the time scales underlying the orbital period variations but cannot use the errors obtained from frequency search as a measure for the accuracy of the obtained values.

\begin{table}
\tabcolsep 2.65mm
\caption{Timescales and amplitudes of orbital period variation.}
\label{Tab06}
\begin{tabular}{lrrrrrr}
\toprule
$P$ (y)  & 6.32 & 6.94 & 8.60 & 10.78 & 14.55 & 52.70\\
$A$ (sec)& 0.34 & 0.37 & 0.37 &  0.24 &  0.26 &  0.45\\
\bottomrule
\end{tabular}
\end{table}

\section{Radial velocity variations with orbital phase}
\subsection{Keplerian approach}\label{Sect7.1}

Figure\,\ref{Fig13} shows the RVs folded with the orbital period after subtracting the best-fitting Keplerian orbits computed with the method of differential corrections as mentioned in Sect.\,\ref{Sect6}. We use the observations from 2006 when RZ\,Cas was in a quiet state for comparison, shown by black dots. The deviations from a straight line (pure Keplerian motion) of the RVs of the secondary seen in 2006 are due to its non-spherical shape and inhomogeneous surface intensity distribution as discussed in \citet{2009A&A...504..991T}, which we investigate in detail in Sect.\,\ref{Sect7.2}. More or less strong deviations from the behaviour in 2006 can be seen in different years. 

Looking at the behaviour of the RVs of the secondary, we find the strongest deviations in 2001 where we see large variations with orbital phase. Almost no differences to 2006 are observed in 2009, 2015, and 2016, whereas moderate differences occur in 2013 and 2014. Interpreting the strength of the deviations as activity indicator, we conclude that we observed RZ\,Cas in or just after a mass-transfer episode in 2001 and in a quiet state in 2006 (as already stated in \citet{2008A&A...480..247L}, and \citet{2009A&A...504..991T}), and that this quiet phase continued until 2009, followed by a slightly more active phase around 2013 and 2014, and falling back into a quiet state in 2015 and 2016.

Looking at the behaviour of the RVs of the primary, in particular at the amplitude and shape of the RME, we see a different picture. The amplitude of the RME in 2001 is much larger and its shape is strongly asymmetric. In all the other years, however, we see almost no difference to 2006, except for three nights of observation covering the ingress of the eclipses: one in 2009, one in 2013, and one in 2015. 

\subsection{Analysis with PHOEBE}\label{Sect7.2}

RZ Cas is a semi-detached Algol-type binary system and its cool component fills its critical Roche lobe. Tidal distortions occur and lead to non-spherical shapes. According to \citet{2013MNRAS.431.2024S}, non-negligible effects on RVs occur for $a<20\,(R_1+R_2)$, that is when the separation of the components is smaller than 20 times the sum of their radii. Thus, the a priori assumption of Keplerian orbit, which considers the mass centre of the stellar disc coinciding with its intensity centre, is no longer valid. As already mentioned in previous sections, RZ Cas is undergoing stages of active mass transfer. A hot region around the equatorial belt has long been known \citep{1982ApJ...259..702O}. Moreover, the recent comprehensive tomographic study of \citet{2014ApJ...795..160R} revealed clear indicators of mass stream activity in several short period Algols, including RZ Cas. \citet{1994PASJ...46..613U} found that the secondary of RZ\,Cas can be modelled only when assuming an unusually large value of the gravity darkening exponent of 0.53 and inferred that dark spots are present on the front and back sides of the secondary with respect to the primary. These authors supposed that quasi-radial flow in the sub-adiabatic stellar envelope from the deep interior is the cause of darkening. \citet{2008ysc..conf...33T, 2009A&A...504..991T} confirmed the finding of the two spots, resuming the interpretation. \citet{2018MNRAS.481.5660D} and \citet{2018A&A...611A..69B}, on the other hand, give an alternative explanation, showing for the short-period Algols \object{$\delta$~Lib} and \object{$\lambda$~Tau} that the mass stream can produce a light scattering cloud in front of the surface of the Roche lobe filling secondary facing the primary.

\subsubsection{Method}

To account for all these effects, we need to use sophisticated models to simulate RV changes during the whole orbital motion via integrating both components surface intensities. For this purpose, we used the well-known Wilson-Devinney code \citep{1971ApJ...166..605W, 2005Ap&SS.296..121V} through the PHOEBE interface \citep{2005ApJ...628..426P}. The WD code consists of light and/or RV curve synthesiser (LC) and parameter optimiser (DC) for fitting purpose. As  we describe later, our attempts to optimise the multi-parameter fit failed with DC, which is expected for such complex configurations. Thus we used LC via our Markov Chain Monte Carlo (MCMC hereafter) optimiser written in Phoebe-scripter extension, to both optimise the parameters and explore the parameter space \citep[see][]{2018MNRAS.481.5660D}.

In MCMC runs, we start from a random initial parameter set and add accepted parameters to the Markov chain. We need to run a lot of chains (or "walkers") to avoid the issue of one Markov chain sticking in a local $\chi^2$ valley. That is why it is advisable to start the simulation with at least two times the number of walkers of the number of parameters to be optimised \citep[see][]{2013PASP..125..306F}. For our last sets of simulation, we set the number of walkers > 20 and the lowest iteration numbers are dynamically increased to fulfil the condition by \citet{2013PASP..125..306F} that the iteration number should be at least ten times the auto-correlation time (i.e. typically 500,000 iterations per season). 

\subsubsection{Application}

Spots are described in WD by two coordinates, co-latitude $\delta$ and longitude $\lambda$, and two physical parameters, temperature ratio compared to the normal surface $T$ and radius $R$. In our initial run, we started to model the RVs of the cool component by taking one spot (Spot1) on the cool secondary facing the primary into account to mimic a scattering cloud or diffusive material between the components. All of our models converged to spot position $\delta\approx 90^\circ$, $\lambda\approx 0^\circ$, that is the region that faces the primary. However, we found a strong correlation between spot size and temperature ratio of the form of $R^2T^4$\,=\,constant, balancing a lower temperature ratio by a more extended spot. Thus, we could only fit the "strength" or "contrast" of the spot with respect to the stellar surface, fixing the temperature ratio to a reasonable lower limit of 0.76. As mentioned at the beginning of Sect.\,\ref{Sect7.2}, \citet{1994PASJ...46..613U} suggested dark spots at the front and back sides of the cool secondary towards the primary caused by mass transfer. We therefore implemented a second spot (Spot2) on the opposite side of the surface of the secondary to check if this offers further improvement.

To check if we can detect a variation of the filling factor between active and inactive phases of RZ\,Cas, we used the "unconstrained mode " of WD and the surface potential of the secondary as a free parameter. In all of our runs, however, the filling factor of the cool component converged to unity. So for the final analysis, we fixed it to unity and used the "semi-detached mode". Eccentricity was set to zero and the orbital inclination to 82$^\circ$. The temperature ratio was fixed to 0.76 for both spots on the secondary and, as already mentioned, the position of Spot1 to $\delta_1$\,=\,90$^\circ$, $\lambda_1$\,=\,0$^\circ$. The location of Spot2 converged to $\delta_2$\,=\,90$^\circ$ without showing larger scatter, so we fixed $\delta_2$ as well.

\subsubsection{Results}\label{Sect7.2.3}

Free parameters were the synchronisation parameter for the primary, $F_1$, the radii of Spot1 and Spot2 on the secondary, $R_1^{sec}$ and $R_2^{sec}$, the longitude of Spot2, $\lambda_2^{sec}$, and the systemic velocity $V_\gamma$. The RV semi-amplitudes were allowed to vary within the error bars derived in previous attempts and led to the values as listed in Table\,\ref{Tab07}. 

\begin{table}\centering
\tabcolsep 4.1mm
\caption{Basic absolute values derived with PHOEBE.} 
    \label{Tab07}
    \begin{tabular}{llll}
    \toprule  
    $q$            & 0.350782(47) & $M_1$ (M$_\sun$) & 1.9507(54)\\
    $a$ (R$_\sun$) & 6.5464(54)   & $M_2$ (M$_\sun$) & 0.6843(13) \\
    \bottomrule
    \end{tabular}
\end{table}

\begin{table}
\tabcolsep 1.45mm
\caption{Synchronisation factor of the primary, radii of both spots on the secondary, longitude of Spot2 on the secondary, and mean scatter of the RV residuals.}\label{Tab08}
\begin{tabular}{lccccc}
\toprule
Season & $F_1$ & $R_1^{sec}$ & $R_2^{sec}$ & $\lambda_2^{sec}$ & rms\\
       &       & \small (deg)& \small (deg)& \small (deg)      & \small (\kms)\\
\midrule
2001  & 1.217(20) & $23.09_{-1.40}^{+0.80}$  & $22.8_{-2.2}^{+2.1}$ & $136.7_{-4.1}^{+3.9}$& 5.31\vspace{1mm}\\
2006  & 0.924(13) & $41.04_{-0.92}^{+0.92}$  & $10.6_{-2.5}^{+1.9}$ & $150  _{-20 }^{+26 }$& 2.47\vspace{1mm}\\
2009  & 1.006(11) & $36.55_{-0.74}^{+0.74}$  & $10.8_{-1.7}^{+1.4}$ & $244.4_{-6.1}^{+6.6}$& 1.62\vspace{1mm}\\ 
2013a & 0.897(08) & $28.69_{-1.08}^{+0.64}$  & $24.0_{-1.0}^{+0.9}$ & $180.0_{-2.4}^{+2.4}$& 3.46\vspace{1mm}\\
2013b & 0.991(14) & $28.01_{-1.09}^{+0.99}$  & $23.7_{-1.2}^{+1.2}$ & $218.4_{-2.9}^{+3.1}$& 3.60\vspace{1mm}\\  
2014  & 0.812(12) & $33.16_{-0.84}^{+0.96}$  & $21.3_{-1.2}^{+1.2}$ & $209.4_{-3.1}^{+3.3}$& 2.92\vspace{1mm}\\  
2015a & 1.057(13) & $42.51_{-0.72}^{+0.66}$  & $19.2_{-1.9}^{+1.8}$ & $194.4_{-3.2}^{+3.6}$& 2.71\vspace{1mm}\\  
2015b & 0.782(11) & $41.53_{-0.54}^{+0.69}$  & $14.3_{-1.0}^{+0.9}$ & $180.0_{-3.0}^{+3.2}$& 2.46\vspace{1mm}\\
2016  & 0.826(10) & $42.71_{-0.73}^{+0.63}$  & $16.2_{-1.1}^{+1.1}$ & $124.5_{-3.3}^{+3.0}$& 2.26\\ 
\bottomrule
\end{tabular}
\end{table}

\begin{figure}\centering
\includegraphics[width=\linewidth]{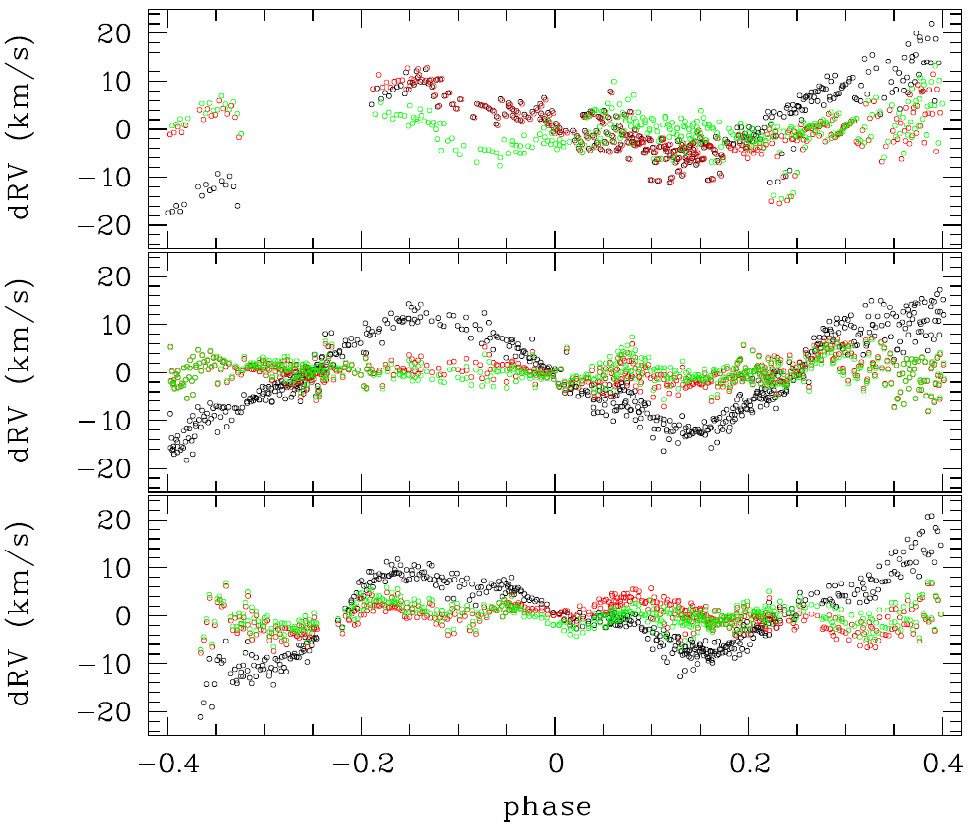}
\caption{Residuals of the RVs of the secondary after subtracting the best-fitting PHOEBE solutions without including spots (black), including one spot (red), and including two spots (green) on the secondary, obtained (from top to bottom) for seasons 2013a, 2015b, and 2016.}\label{Fig14}
\end{figure}

\begin{table}\centering
\caption{Absolute values of positive and negative RV deviations due to RME.}
\begin{tabular}{lccc}
    \toprule
Season& JD & $K_p$ (\kms)& $K_n$ (\kms)\\
    \midrule
2001  & 2\,452\,192 & 26.5 & 32.4\\
2006  & 2\,453\,768 & 19.9 & 22.4\\
2009  & 2\,455\,156 & 24.7 & 21.0\\
2013a & 2\,456\,584 & 22.6 & 21.7\\
2013b & 2\,456\,672 & 26.4 & 20.0\\
2014  & 2\,456\,936 & 19.3 & 19.7\\ 
2015a & 2\,457\,238 & 24.6 & 19.8\\
2015b & 2\,457\,293 & 17.5 & 19.8\\
2016  & 2\,457\,646 & 19.1 & 19.0\\
\bottomrule
    \end{tabular}
    \label{Tab09}
\end{table}

In Fig.\,\ref{FigA1}, we show the corner plot for season 2009. This shows that MCMC not only gives the most optimal parameter set, it also provides an error estimation on each parameter and possible correlations between the parameters. Finally obtained values are listed in Table\,\ref{Tab08}, in which we also included the mean scatter calculated from the RV residuals after subtracting the PHOEBE solutions obtained for each season. These residuals are shown in Fig.\,\ref{FigA2}. 

Figure\,\ref{Fig14} shows the influence of the inclusion of spots on the secondary into the model. We selected three seasons as examples. Including spots on the secondary did not effect the modelled RVs of the primary and vice versa. Thus, we only show the residuals of the RVs of the secondary. The black dots corresponds to the best-fitting solutions without spots, that is when only the non-spherical shapes of the components (in particular of the secondary) are taken into account. The RVs show a systematic deviation around Min\,II that is distinctly reduced when including Spot1 that faces the primary, resulting in the red dots. The RVs show smaller, non-systematic deviations around Min\,I, which is clearly present in the data from 2013a, small in 2016, and almost not visible in 2015b. Adding the second spot on the opposite side reduces this scatter, as shown by the green dots, but in some cases the reduction is marginal.

One explanation for the asymmetry observed in the RME in the years 2009, 2013b, and 2015a (see Fig.\,\ref{Fig13}) could be the impact of a hot spot on the primary on its RVs. For comparison, we therefore additionally included a hot spot on the primary for
these seasons, taking three more free parameters (temperature ratio, radius, and longitude) into account. The fit did not improve the solutions, however. 

\begin{figure}
\includegraphics[width=\linewidth]{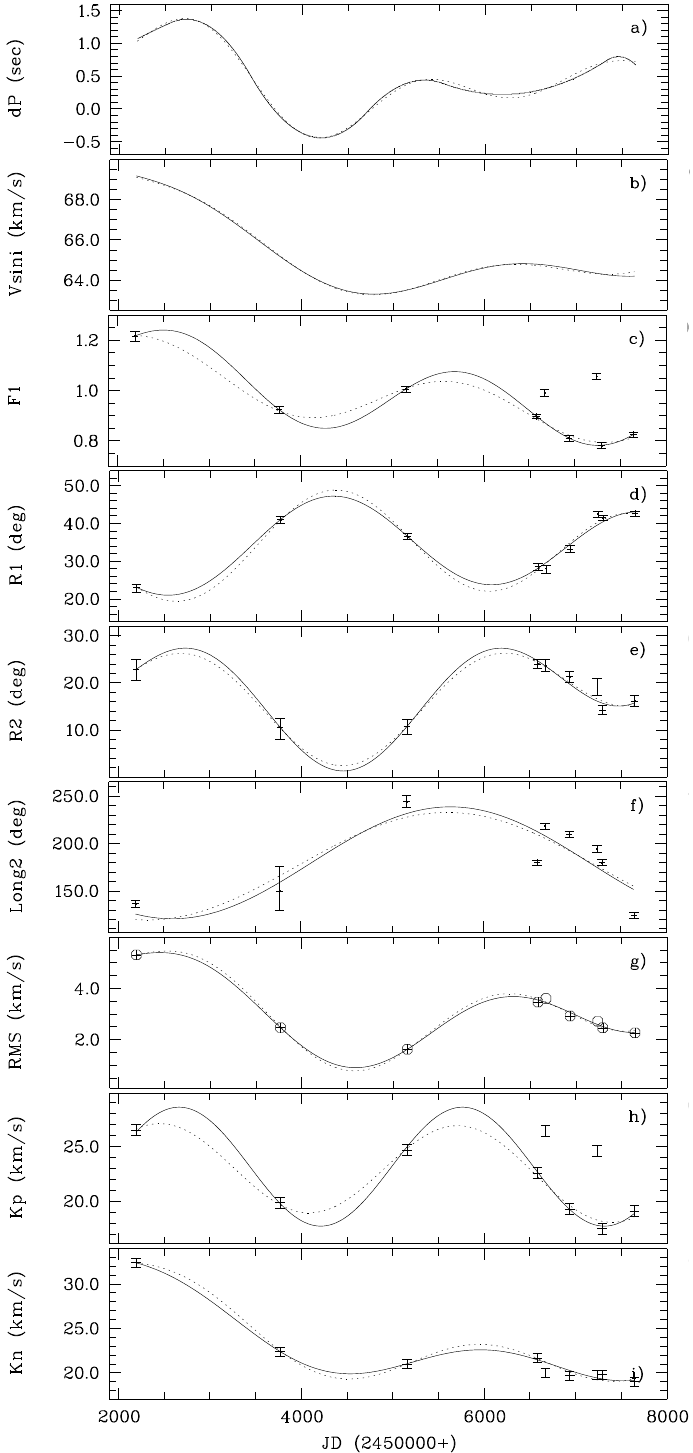}
\caption{Time variations of different parameters (see text). The solid lines are calculated from the best-fitting periods and long-term trends. The dotted lines show possible sinusoidal variations with a period of 9\,yr plus a long-term trend, except for panel f), which shows an 18 y variation. Values from 2013b and 2015a were considered as outliers.}\label{Fig15}
\end{figure}

Table\,\ref{Tab09} lists absolute values of positive and negative amplitudes of the RME (maximum deviations in RV during ingress and egress, respectively) obtained after subtracting the best-fitting PHOEBE solution for $F_1$\,=\,0, that is for a non-rotating primary, from the observed RVs. Panels h) and i) in Fig.\,\ref{Fig15} show that the positive amplitudes reveal a similar behaviour as $F_1$ or $R_2$, whereas the negative amplitudes show a systematically decreasing trend with time (see next Section for a discussion). 

\begin{table}\centering
\caption{Timescales in years determined from the seasonal variations of different parameters.}
\begin{tabular}{ccccccccc}
\toprule
d$P$ & \vs & $F_1$ & $R_1^{sec}$ & $R_2^{sec}$ & $\lambda_2^{sec}$ & rms & $K_p$ & $K_n$\\
\midrule
8.6 & 9.6 & 8.4 & 9.4 & 8.8 & 16.7 & 9.3 & 8.5 & 9.6\\
\bottomrule
\end{tabular}\label{Tab10}
\end{table}

Figure\,\ref{Fig15} shows the seasonal variations of all investigated parameters. In panel a) we added the period change derived from the O-C values (cf. Sect.\,\ref{Sect6}), in panel b) the \vs\ values from the fit determined in  Sect.\,\ref{Sect5.4}, and in panel g) the mean scatter of the RV residuals after subtracting the best-fitting PHOEBE solutions from the input data (see Fig.\,\ref{FigA2}). All variations can be described by a sinusoid plus a long-term trend. Because the seasonal sampling was not sufficient to determine a second frequency describing the long-term trend, we fixed it to $10^{-6}$\,\cd\ and determined only amplitudes and phases. Table\,\ref{Tab10} lists the timescales derived from the best-fitting sinusoids.

The best-fitting curves are shown in Fig.\,\ref{Fig15} by solid lines. We see that $F_1$, $R_2$, rms, and $K_p$ vary almost in phase, whereas $R_1$ varies in anti-phase. Moreover, \vs\ and $K_p$ show almost identical shapes of variation. Searching for a possible common period that explains the variations of all parameters, we found a period of 9 years that best fits, together with the mentioned long-term trend, the variations of all parameters except for $\lambda_2^{sec}$ that can be fitted by a period of 18\,years, twice the period of 9\,years. For d$P$ we had to add a third (optimised) period of 6.3\,yr. The resulting fits are shown in Fig.\,\ref{Fig15} by  dotted lines. In all cases, the quality is comparable to that obtained from the optimised values listed in Table\,\ref{Tab10}.

\section{Discussion}\label{Sect8}

\begin{figure}
\includegraphics[angle=-90, width=\linewidth]{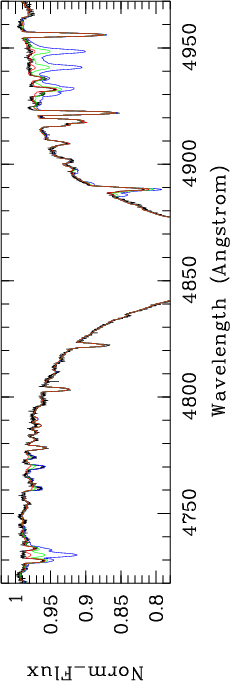}
\caption{Fit of the observed spectrum of the primary (black) in the H$_\beta$\,region by synthetic spectra with [C/H] of 0.0 (blue), $-0.4$ (green), and $-0.8$ (red).}
\label{Fig16}
\end{figure}

Our spectral analysis yields atmospheric parameters of the components that are in agreement with the results of light curve analysis by \citet{2004MNRAS.347.1317R}. For the first time, we determine the elemental surface abundances. For both components, we find [Fe/H] close to $-0.4$, as for Ca, Cr, Mn, and Ni of the primary, whereas O, Mg, Si, Sc, Ti, and V show relative abundances between $-0.2$ and $-0.1$. For the carbon abundance of the primary, we find [C/H]\,=\,$-0.80^{+0.13}_{-0.18}$, which is remarkably different from the other elements. We think that the difference is significant. First, the considered spectral range shows a sufficiently large number of C\,I lines of the primary (SynthV computes line depths >5\% for 32 rotationally unbroadened C\,I lines). Second, differences between the fits with [C/H] of $-0.4$ and $-0.8$ can clearly be seen by eye. Figure\,\ref{Fig16} shows this for the strongest carbon lines in the H$\beta$ region of the spectrum of the primary (observed composite spectrum after subtracting the synthetic spectrum of the secondary). In the spectrum of the cool secondary, carbon is mainly present in form of the CH molecule bands. Compared to the primary, the signal is very weak and we can only say that [C/H] is below solar abundance, between $-0.3$ and $-1.0$ within the 1$\sigma$ error bars. 

\citet{2008A&A...486..919M} tried to model the binary evolution of RZ\,Cas and found consistent solutions only for an initial mass ratio of $q$\,$\approx$\,3, which is about the inverse of the actual ratio. From the large initial mass of the donor in that case we can assume that the primary has switched during its evolution from pp-chain to CNO-cycle hydrogen burning, resulting in depleted carbon and enhanced nitrogen abundances in its core \citep[e.g.][]{2010A&A...517A..38P}. From the reversal of mass ratio we conclude that the donor was stripped by mass loss down to its core so that the gainer accreted  CNO-cycled, carbon-deficient material in the late phase of its evolution. This material has then mixed with the surface layers of the gainer, leading to the observed carbon abundance. In that case, the surface carbon abundance of the gainer cannot be lower than that of the donor star, however. Multiple authors have used the sketched scenario to explain the surface abundance anomalies observed in several other Algol-type stars; these include \citet{2012MNRAS.419.1472I} for \object{GT~Cep}, \object{AU~Mon}, and \object{TU~Mon}, \citet{2014MNRAS.444.3118K} for \object{u~Her}, and \citet{2018MNRAS.481.5660D} for $\delta$\,Lib.

\begin{figure*}
\includegraphics[width=.5\linewidth]{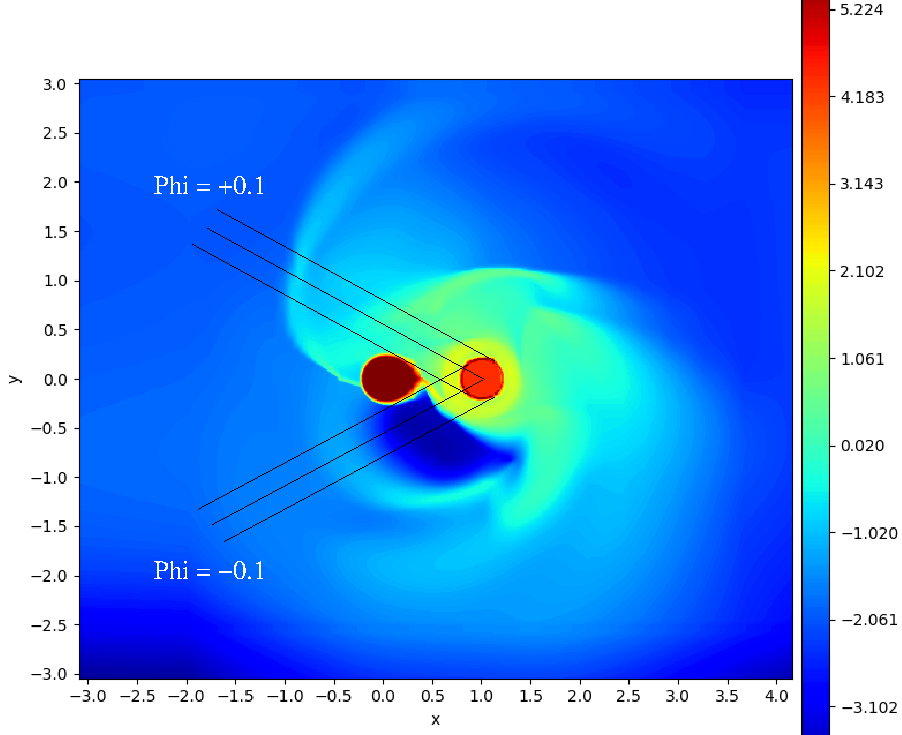}
\includegraphics[width=.5\linewidth]{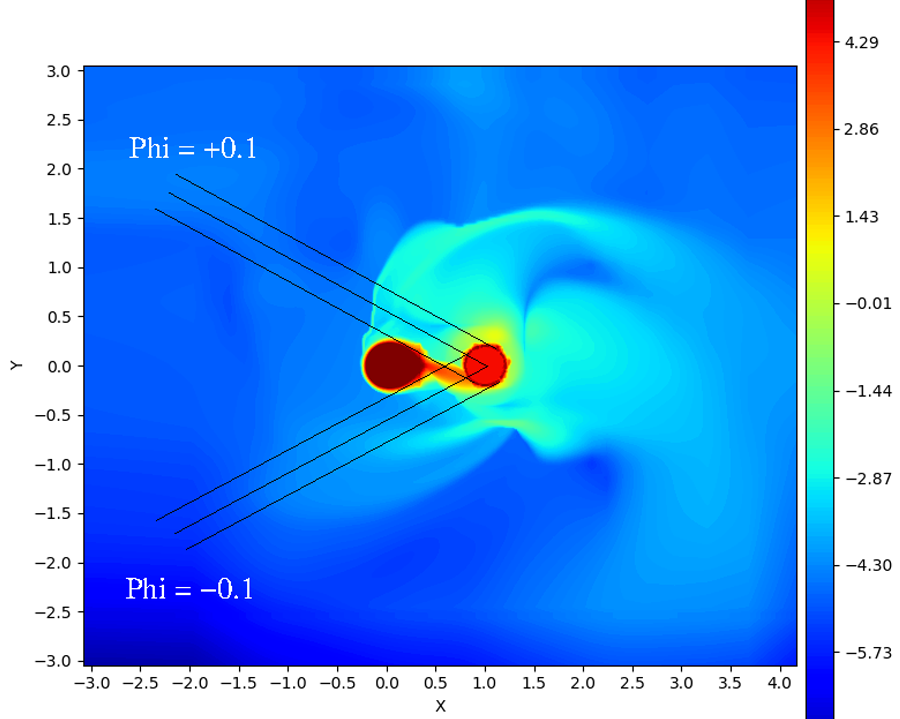}
\caption{Logarithmic plots of gas density distributions (in $10^{11}$\,cm$^{-3}$) obtained from 3D hydrodynamic simulations  of the RZ\,Cas system (Nazarenko \& Mkrtichian, priv. comm.)  for  mass-transfer rates of 1\,$\times$\,$10^{-9}$\,M$_\odot$y$^{-1}$ (left) and 6\,$\times$\,$10^{-8}$\,M$_\odot$y$^{-1}$ (right). The viewing angles at first ($\phi$\,=\,$-0.1$) and last ($\phi$\,=\,$+0.1$) contact are indicated.}
\label{Fig17}
\end{figure*}

Our further investigation was based on the separated LSD profiles and RVs of the two components of RZ\,Cas computed with LSDbinary, and on the times of minimum (O-C values) taken from literature. The change in orbital period computed from the slope of the O-C diagram cannot be characterised by one single timescale, as already found by \citet{2018MNRAS.475.4745M}. These authors derived timescales of 4.8, 6.1, and 9.2 years. We observe from our slightly extended data set different timescales of variation in different seasons, such as 6.3\,yr around JD\,2\,440\,000, and 8.6\,yr from 2\,450\,000 to 2\,455\,000 (which is about the time span of our spectroscopic observations), overlaid by longer periods. From our previous analysis using the Shellspec07\_inverse code \citep{2009A&A...504..991T}, we know that RZ\,Cas was in an active state of mass transfer in 2001 and in a quiet state in 2006. Comparison of the RV residuals after subtracting a pure Keplerian orbit with those from 2006 gave us initial hints pointing to further activity periods (Fig.\,\ref{Fig13}). From the RVs of the primary we see that the RME was distinctly enhanced only in 2001. A comparison of the RV residuals of the secondary with those from 2006 shows the largest deviation in 2001, weaker deviations in 2013 and 2014, and almost no deviations in 2009, 2015, and 2016. The amount of scatter found in the RV residuals after subtracting the best-fitting PHOEBE solutions is strongly correlated with this finding. Thus, we conclude that no further mass-transfer episode as strong as in or close to 2001 occurred in RZ\,Cas.

The modelling of the observed RVs of both components with MCMC-PHOEBE gave the best results when adding two dark spots on the surface of the cool companion: Spot1 facing the primary, Spot2 on the opposite side.  Spot1 was already found from the RVs from 2001 and 2006 by \citet{2009A&A...504..991T}, whereas \citet{1994PASJ...46..613U} included two spots into their model, explaining the observed  anomalous gravity darkening by a cooling mechanism by enthalpy transport due to mass outflow that leads to a reverse process of gravitational contraction. 

We monitored both spots for the first time over decades.
We find that Spot1 always exactly points towards the primary, with radii (a synonym for strength or contrast as mentioned in before) between 23$^\circ$ and 43$^\circ$. Spot2, on the other hand, is much weaker, shows a variation in its position, and induces only a second order improvement in some of the observed seasons (cf. Fig.\,\ref{Fig14}). The main findings are that the strengths of the two spots vary in anti-phase and that Spot2 shows different positions in longitude, varying around the longitude of the L2 point of 180$^\circ$.

The fact that the strength of Spot1 is largest when the star is between 2006 and 2009 in a quiet phase (Fig.\,\ref{Fig15}) speaks against the cooling by mass-outflow mechanism. Instead, we argue that the variations of the strengths of both spots can be explained by magnetic activity of the cool companion, assuming an activity cycle of 9 years, based on an 18-year cycle of magnetic field change, including a reversal of the magnetic poles. The 9-year cycle can be found in the variations of all investigated parameters except for the longitude of Spot2, as we showed in the last section, and was also found by \citet{2018MNRAS.475.4745M}. Spot2 shows a migration in longitude, returning after an 18-year cycle to its position from 2001 (cf. Fig.\,\ref{Fig15}). We assume that we observe similar surface structures on the cool secondary of RZ\,Cas such as for cool, rapidly rotating RS\,CVn binaries or single rapidly rotating variables of FK Com and BY Dra type. Long-living active regions were observed on opposite sides of these stars. The spots are of different intensity and also show switching of activity from one spot to the other on timescales of years or decades, which is known as the flip-flop effect \citep[see][and references herein]{2018A&A...613A...7Y}.

We believe that there is a  direct analogy with sunspot activity. Sunspots show a saucer-shaped depression in the photosphere caused by the Lorentz force of the strong magnetic field within the spot, the so-called Wilson depression \citep{1774RSPT...64....1W}. This means that inside the spot the level of $\tau$\,=\,1 is located below the level of the photosphere outside the spot. For the Sun, the geometric depth of the depression is of the order of 600\,km \citep[e.g.][]{1972SoPh...26...52G, 2018A&A...619A..42L}. For the Roche-lobe filling donor, the existence of such atmospheric depression close to the L1 point means that L1 is "fed" by atmospheric layers of lower density and thus the mass transfer is the more suppressed the deeper the depression is. The local magnetic field strength controls the strength (depression) of Spot1 and the height of atmospheric layers feeding the L1 point and in this way the mass-transfer rate. RZ\,Cas showed, in perfect agreement with the drafted scenario, high mass-transfer rate in 2001 (and possibly in 2011, when observations are missing) when the Spot1 size was around 20 degrees and low activity state in 2006-2007 and in 2015-2016 when the spot size was about 40 degrees.

The explanation for the asymmetry and different amplitudes observed in the RME in the years 2001, 2009, 2013b, and 2015a (see Fig.\,\ref{Fig13}) could be the impact of the combined effect of acceleration of the photospheric layers in 2001 caused by the high mass-transfer rate and by screening the surface of the primary by the dense gas stream \citep{2008A&A...480..247L}. The amplitudes of the RME in RZ\,Cas in general show very different behaviour during ingress ($K_p$) and egress ($K_n$) of primary eclipse. The variation of $K_p$ is similar to those of $R_2$ and rms, whereas the shape of the $K_n$ variation perfectly matches that of \vs\ (see Fig.\,\ref{Fig15}). We assume that $K_n$ is related to the true \vs\ seen outside the eclipses and $K_p$ is strongly influenced by mass-transfer effects. This can be explained from Fig.\,\ref{Fig17}, showing the gas density distribution around RZ\,Cas for two different mass-transfer rates, calculated from 3D-hydrodynamic simulations by Nazarenko \& Mkrtichian (priv. comm.). It can be seen that the equatorial zone of the surface of the primary is masked during ingress (at orbital phases $\phi$\,=\,$-0.1..0$) by the optical thick gas stream from the secondary, whereas we directly see the complete disc of the primary during egress ($\phi$\,=\,$0..0.1$), only very slightly hampered by optical thin circumbinary matter. In consequence, $K_p$ is affected by the seasonal varying density of the gas stream (or mass-transfer rate) and $K_n$ is correlated with the surface rotation velocity of the primary. All the variability seen in RME in the years from 2006 on is mainly related to the variations of $K_p$, forced by changes in gas-stream density and variable screening and attenuation effects, while the rotation speed was nearly constant. 

The synchronisation factor derived with PHOEBE is mainly based on the shape of the RME and this shape is strongly influenced by timely varying Algol-typical effects (see paragraph below). Our MCMC simulation did not count for the observed asymmetry in the RME either (i.e. the fact that the amplitudes $K_p$ and $K_n$ are different from each other) and so the synchronisation factor $F_1$ has to be considered as some kind of mean value. The shape of its time variation resembles the mean of the shapes of the $K_p$ and $K_n$ variations (cf. Fig.\,\ref{Fig15}). The results obtained for the synchronisation factor in Sect.\,\ref{Sect5.4} from radius and \vs\ of the primary, on the other hand, give sub-synchronous rotation of the primary. Its outer layers are only accelerated to almost synchronous rotation during the active phase
in 2001. The finding of sub-synchronous rotation is surprising. It was suggested by \citet{2014A&A...570A..25D} to occur during the rapid phase of mass exchange in Algols, when the donor star is spun down on a timescale shorter than the tidal synchronisation timescale and material leaving the inner-Lagrangian point is accreted back onto the donor, enhancing orbital shrinkage. But these authors state that once the mass-transfer rate decelerates and convection develops in the surface layers, tides should be effective enough to re-synchronise the primary. To our knowledge, sub-synchronous rotation was found so far in only one short-period Algol, namely TV\,Cas, by \cite{1992A&A...257..199K}, in which the authors found synchronisation factors of 0.85 for the primary and 0.65 for the secondary component.

The RV residuals after subtracting the solutions obtained with our MCMC simulations (Fig.\,\ref{FigA2}) finally show that our model distinctly reduces the scatter compared to the residuals obtained from pure Keplerian motion. On the other hand, we can still recognise many of the signs of activity that we discussed in connection with Fig.\,\ref{Fig13}. We assume that  these still unexplained features are due to the  distribution and density of circumbinary matter along the line of sight in different orbital phases, varying between different seasons according to the varying activity of the star. Finally, we can conclude from Fig.\,\ref{FigA2} in the same way as from Fig.\,\ref{Fig13} that RZ\,Cas showed an extraordinary phase of activity around 2001 followed by a quiet phase in 2006 to 2009, slightly enhanced activity in 2013/2014, followed by a quiet phase again. The calculated fits on the variation of the radii of the two spots and the RME amplitude $K_p$ based on the nine-year cycle, on the other hand, show extrema of the same amplitude such as in 2001 for the time around 2010-2011. Therefore, because of missing data in this period, we cannot exclude the possibility that a second mass-transfer episode of comparable strength like in 2001 occurred close to 2010-2011.

The orbital period was at maximum shortly after 2001, dropped down steeply until 2006, and was then more or less continuously rising. From Eq.\,8 in \citet{1973A&A....27..249B}, assuming conservative mass transfer and conservation of orbital angular momentum, we obtain a mass-transfer rate of 1.5\,$\times\,10^{-6}$\,M$_\odot$\,yr$^{-1}$. This is a typical value observed for Algol-type stars \citep[e.g.][]{1976IAUS...73..283H} and also agrees \citet{1976AcA....26..239H}, who calculate, from an O-C analysis, a mean mass-transfer rate of RZ\,Cas of 1.0\,$\times\,10^{-6}$\,M$_\odot$\,yr$^{-1}$, averaged over several episodes of period change. As mentioned in the introduction, evolutionary scenarios for RZ\,Cas have to count for mass loss from the system \citep{2008A&A...486..919M}, which may be the case for Algols in earlier stages of their evolution in general \citep[e.g.][]{2006MNRAS.373..435I, 2013A&A...557A..40D, 2015A&A...577A..55D}. Thus, conservatism is not necessarily justified and the assumption of mass conservation can only lead to a raw estimation of the amount of mass transferred during the active phase of RZ\,Cas in 2001.

We assume that the enhanced value of \vs\ in 2001 points to an acceleration of the outer layers of the primary of RZ\,Cas due to the mass-transfer episode occurring close to that year. This then dropped down by about 5\,\kms\ and it could be that its slight increase after 2009 is correlated with a slightly increased activity as indicated by the variation of $F_1$ and rms.

\section{Conclusions}\label{Sect9}

In this first of three articles related to spectroscopic long-term monitoring of RZ\,Cas, we investigated high-resolution spectra with respect to stellar and system parameters based on the RVs and LSD profiles of its components calculated with the LSDbinary program. The main goal was to search for further episodes of enhanced mass transfer occurring after 2001 and for a general timescale of variations possibly caused by the magnetic cycle of the cool companion.

From spectrum analysis we determined precise atmospheric parameters, among them low metal surface abundances, in particular [Fe/H] of  $-0.42$ and [C/H] of $-0.80$ for the primary and [Fe/H] of the same order for the secondary. The carbon deficiency observed for the primary gives evidence that the outer layers of the cool secondary have been stripped in the fast mass-transfer phase down to its core so that CNO-cycled material was transferred to the outer layers of the primary in later stages of evolution. The derived \te\ of the components and the \lg\ of the primary agree within the 1$\sigma$ error bars with the results from LC analysis by \citet{2004MNRAS.347.1317R}. From the RV analysis with PHOEBE, we derived very precise masses and separation of the components of $M_1$\,=\,1.951(5)\,M$_\odot$, $M_2$\,=\,0.684(1)\,M$_\odot$, and $a$\,=\,6.546(5)\,R$_\odot$.

From several of the investigated parameters that show seasonal variations, such as orbital period, \vs, strength of the spots on the secondary, synchronisation factor calculated with PHOEBE, and rms of the RV residuals after subtracting the PHOEBE solutions, we deduce a common time scale of the order of 9 years. The variation of the orbital period is complex and can be described in detail only when adding further periods. We conclude that we see the effects of a 9-year magnetic activity cycle of the cool companion of RZ\,Cas, caused by an 18-year dynamo cycle that includes a reversal of polarity. This conclusion is strongly supported by the behaviour of the two dark spots on the surface of the secondary that show the flip-flop effect in their strengths and one  spot that shows an 18-year periodicity in longitudinal migration.

From the variations of orbital period and \vs\ around 2001, we conclude that the determined \vs\ is the projected equatorial rotation velocity of the outer layers of the primary accelerated by transferred matter and does not stand for the rotation velocity of the star as a whole. In all other seasons, the measured \vs\ point to a sub-synchronous rotation of the gainer. At the present stage, we cannot give a physical explanation for this new and interesting finding, however.

Based on our available data, we conclude that RZ\,Cas was undergoing an episode of high mass transfer in 2001, in a quiet phase in 2006 and 2009, followed by a slightly more active phase in 2013 and 2014, and again in a quiet phase in 2015 and 2016. Because we did not observe the star in 2010 and 2011, we cannot exclude that a second episode of high mass transfer occurred in these years, which would agree with the derived magnetic activity cycle of nine years.

The results of our investigation of high-frequency oscillations of the primary of RZ\,Cas will be presented in Part II of this article. In a third article, we will investigate the accretion-induced variability of He\,I lines detected in the spectra.

\begin{acknowledgements}
HL and FP acknowledge support from DFG grant LE 1102/3-1. VT acknowledges support from RFBR grant No. 15-52-12371. AD is financially supported by the Croatian Science Foundation through grant IP 2014-09-8656 and Erciyes University Scientific Research Projects Coordination Unit under grant number MAP-2020-9749. The research leading to these results has (partially) received funding from the European Research Council (ERC) under the European Union's Horizon 2020 research and innovation programme (grant agreement N$^\circ$670519: MAMSIE), from the KU~Leuven Research Council (grant C16/18/005: PARADISE), from the Research Foundation Flanders (FWO) under grant agreement G0H5416N (ERC Runner Up Project), as well as from the BELgian federal Science Policy Office (BELSPO) through PRODEX grant PLATO. Results are partly based on observations obtained with the HERMES spectrograph, which is supported by the Research Foundation - Flanders (FWO), Belgium, the Research Council of KU Leuven, Belgium, the Fonds National de la Recherche Scientifique (F.R.S.-FNRS), Belgium, the Royal Observatory of Belgium, the Observatoire de Gen\`eve, Switzerland and the Th\"uringer Landessternwarte Tautenburg, Germany.
\end{acknowledgements}

\bibliographystyle{aa}
\bibliography{aa39355-20} 

\begin{appendix}
\onecolumn
\section{Analysis of radial velocity variations with PHOEBE}

\begin{figure*}[hbt!]\hspace{-4mm}
\includegraphics[width=1.05\textwidth]{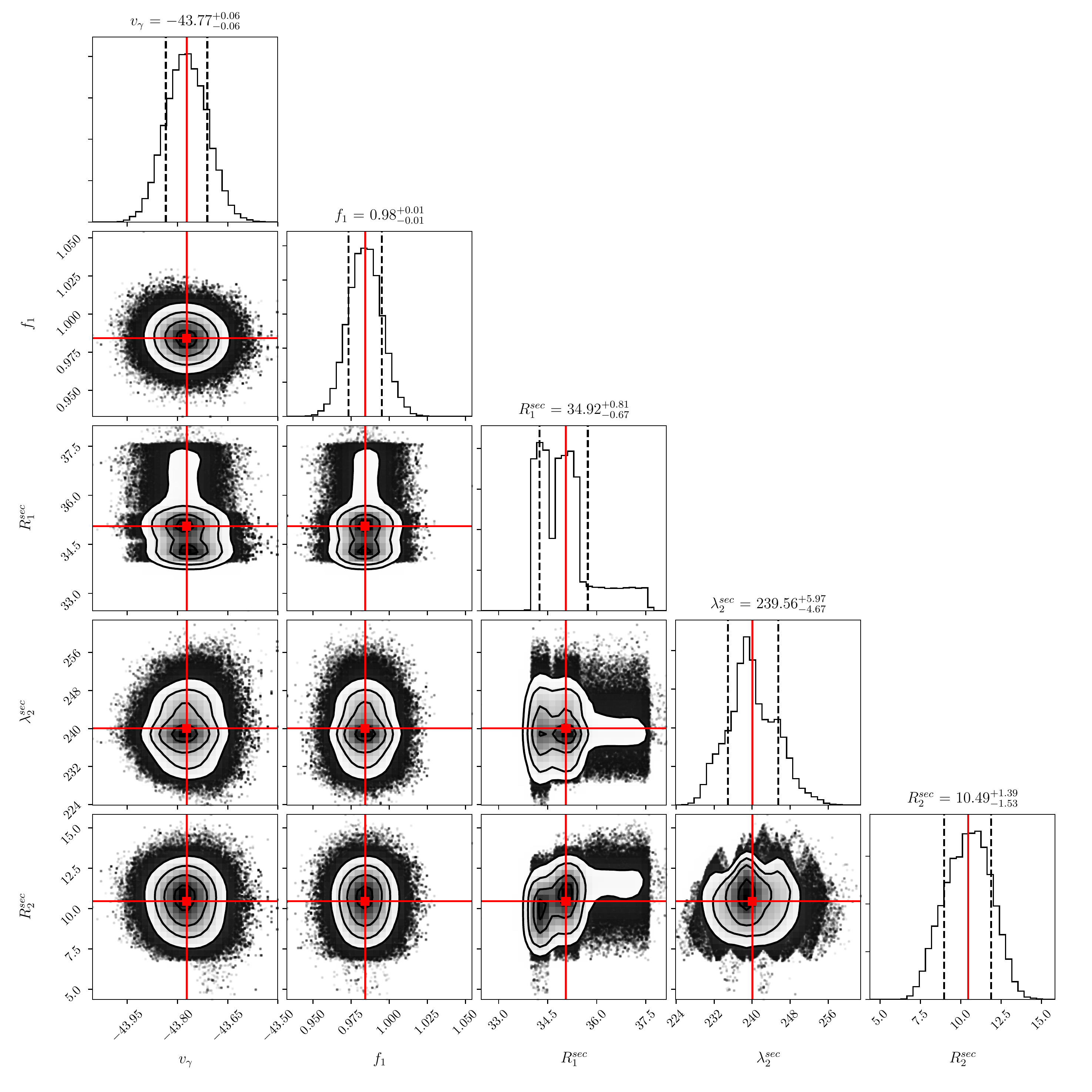}
\caption{Corner plot for the year 2009, showing the $\chi^2$ distributions for systemic velocity, synchronisation factor of the primary, radius of the spot on the secondary facing the primary, and longitude and radius of the opposite spot on the secondary.}
\label{FigA1}
\end{figure*}

\begin{figure*}[hbt!]\centering
\includegraphics[width=\textwidth]{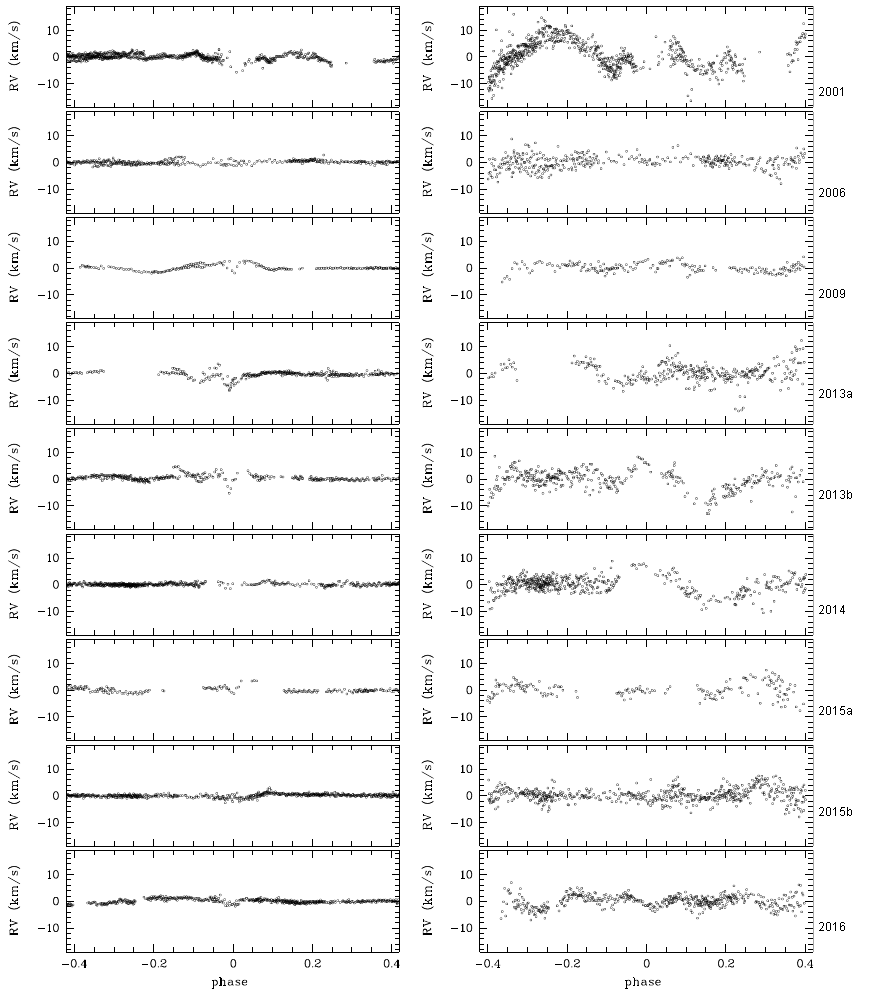}
\caption{Residuals after subtracting the PHOEBE solutions from the RVs of the primary (left) and the secondary (right) component of RZ\,Cas.}
\label{FigA2}
\end{figure*}

Figure\,\ref{FigA1} shows an example of a corner plot obtained with the PHOEBE-based MCMC method. 
Figure\,\ref{FigA2} shows the RV residuals after subtracting the PHOEBE solutions (cf. Sect.\,\ref{Sect7.2.3}).

\end{appendix}
\end{document}